\documentclass[a4paper,11pt]{article}
\usepackage{jcappub} 
\usepackage{lineno}
\usepackage{float}
\usepackage{makecell}
\usepackage{multirow}
\usepackage{hhline}
\usepackage{etoolbox}

\usepackage{tocloft}
\usepackage[title]{appendix}

\makeatletter
\renewcommand\appendix{%
  \par
  \setcounter{section}{0}%
  \setcounter{subsection}{0}%
  \renewcommand\thesection{\Alph{section}}%
  \renewcommand\thetable{\Alph{section}\arabic{table}}%
  \renewcommand\thefigure{\Alph{section}\arabic{figure}}%
  \addtocontents{toc}{%
    \protect\renewcommand{\protect\cftsecpresnum}{Appendix\space}%
    \protect\renewcommand{\protect\cftsecaftersnum}{\space}%
  }%
  \addtocontents{toc}{\protect\setlength{\protect\cftsecnumwidth}{7em}}%
}
\makeatother

\makeatother

\usepackage{booktabs}
\usepackage{graphicx}
\usepackage{subcaption}
\usepackage{amsmath}
\usepackage{array}
\usepackage[margin=1in]{geometry}
\usepackage{adjustbox}

\def\gsim{\mathrel{\raise.3ex\hbox{$>$\kern-.75em\lower1ex\hbox{$\sim$}}}}

\def\lsim{\mathrel{\raise.3ex\hbox{$<$\kern-.75em\lower1ex\hbox{$\sim$}}}}

\title{\boldmath Evaluating the Contribution of Active Galactic Nuclei to the Diffuse High-Energy Neutrino Flux}

\author{Samyak Jain, Dan Hooper, and Francis Halzen}
\affiliation{University of Wisconsin-Madison, Department of Physics and the Wisconsin IceCube Particle Astrophysics Center}

\emailAdd{samyak@icecube.wisc.edu}
\emailAdd{dwhooper@wisc.edu} 
\emailAdd{halzen@icecube.wisc.edu}

\abstract{The detection of high-energy neutrinos from NGC 1068 and TXS-0506+56 suggests that active galactic nuclei (AGN) may contribute significantly to the the diffuse neutrino flux measured by IceCube. Using 10 years of publicly available IceCube data, we performed a systematic population analysis of X-ray-bright and gamma-ray-bright AGN to evaluate the extent to which this diffuse flux could originate from these sources. We find that gamma-ray-bright blazars can account for no more than 16\% of IceCube's total diffuse flux. Although we find no evidence of neutrino emission from gamma-ray-bright, non-blazar AGN, we cannot exclude the possibility that these sources contribute significantly to the diffuse flux. In contrast, we report (pre-trials) evidence of neutrino emission from several nearby, X-ray-bright, Seyfert-type AGN, including \mbox{NGC 1068} ($4.9\sigma$), SWIFT J1041.4-1740 ($2.6\sigma$), SWIFT J0202.4+6824A/B ($2.6\sigma$), SWIFT J0744.0+2914 (2.6$\sigma$), NGC 4151 ($2.5\sigma$), and NGC 3079 ($2.5\sigma$). Although not fully conclusive, these results suggest that IceCube may be detecting neutrinos from a larger population of Seyfert galaxies. The fact that these sources are not gamma-ray bright indicates that their neutrino production must be taking place in optically thick environments, such as in the coronae surrounding these galaxies' supermassive black holes. We also identify a $4.2\sigma$ correlation between the neutrinos detected by IceCube and members of the Swift-BAT catalog of X-ray-bright AGN, although this correlation is dominated by NGC 1068. We estimate that this class of sources contributes between 11.2\% and the entirety of IceCube's total diffuse neutrino flux. These results strengthen the emerging case for the prevalence of gamma-ray-obscured AGN as significant sources of high-energy neutrinos.}

\begin{document}
\maketitle
\flushbottom
\renewcommand{\arraystretch}{1.2} 

\section{Introduction}
In 2013, the IceCube collaboration reported their detection of a diffuse and approximately isotropic flux of high-energy astrophysical neutrinos~\cite{IceCube:2013low}; see also, Refs.~\cite{IceCube:2020acn,IceCube:2021rpz,IceCube:2013cdw,IceCube:2014stg}. Since that time, greater statistics and improvements to the detector have enabled IceCube to characterize the spectral shape of this flux~\cite{icecube_diffuse_recent} and to identify the first sources of these high-energy neutrinos~\cite{txs_icecube_archival,ngc1068_icecube,icecube_ngc4151,Abbasi:2025tas}. These sources include the blazar TXS 0506+56~\cite{IceCube:2018dnn,IceCube:2018cha}, the nearby Seyfert galaxy NGC 1068~\cite{ngc1068_icecube}, and the Milky Way's Galactic Plane~\cite{icecube_galactic_plane}. IceCube has also ruled out many classes of astrophysical objects as bright sources of high-energy neutrinos, including gamma-ray bursts~\cite{grb_no_neutrinos,2012Natur.484..351I,IceCube:2018omy} and gamma-ray-bright blazars~\cite{IceCube:2025tmc,IceCube:2016qvd,Smith:2020oac,Hooper:2018wyk}.

Active Galactic Nuclei (AGN) have long been considered promising as sites of cosmic-ray acceleration~\cite{Biermann:1987ep} and high-energy neutrino production~\cite{agn_neutrino}. AGN are the most luminous persistent sources of electromagnetic radiation in the universe and constitute the majority of all detected gamma-ray sources~\cite{Fermi-LAT:2019yla}. The detection of high-energy neutrinos from \mbox{TXS 0506+56} and NGC 1068 demonstrates that AGN do indeed accelerate cosmic rays, and further raises the question of whether a significant fraction of the diffuse flux observed by IceCube might originate from this class of astrophysical objects. 

NGC 1068 is a nearby ($d=12.7 \, {\rm Mpc}$~\cite{Ricci:2017dhj}) and X-ray-bright ($L_X = 7 \times 10^{43} \, {\rm erg/s}$~\cite{Marinucci:2015fqo,Bauer:2014rla}), Seyfert-type AGN. IceCube's detection of a statistically significant flux of neutrinos with energies of $\sim 1-10 \, {\rm TeV}$ from this source~\cite{ngc1068_icecube}, combined with the strong upper limits on its TeV-scale gamma-ray emission~\cite{MAGIC:2019fvw}, indicates that it must produce neutrinos through pion production in the dense and optically-thick region that surrounds its supermassive black hole~\cite{Blanco:2023dfp,Das:2024vug,Ajello:2023hkh,Murase:2022dog,Blanco:2025zqo}; see also, Refs.~\cite{Saurenhaus:2025ysu,Ambrosone:2024zrf,IceCube:2024ayt,Inoue:2024nap,Fang:2023vdg}. More generally speaking, it is becoming increasingly clear that the majority of the astrophysical neutrinos detected by IceCube do not come from bright sources of gamma rays. Instead, these neutrinos must originate either from a large population of relatively faint sources, such as starforming galaxies~\cite{Loeb:2006tw,Lunardini:2019zcf}, for example, or from obscured or ``hidden'' sources from which gamma rays cannot escape~\cite{francis_gamma_ray_obscured,Hooper:2016jls,Murase:2015xka,Giacinti:2015pya,Murase:2013rfa,IceCube:2023htm,IceCube:2019cia,IceCube:2016tpw,IceCube:2018omy,IceCube:2016ipa,Smith:2020oac,IceCube:2016qvd,Hooper:2018wyk}. In this context, the dense cores of X-ray-bright active galaxies are a particularly promising class of potential sources~\cite{Murase:2015xka,Khiali:2015tfa,Stecker:2013fxa,Kimura:2014jba,Kalashev:2014vya}. The results of this study reinforce the emerging case for gamma-ray obscured AGN as a major contributor to the diffuse high-energy neutrino flux.


In this study, we use 10 years of publicly available IceCube data~\cite{DVN/VKL316_2024} to search for high-energy neutrinos from various populations of blazar and non-blazar AGN. This work follows similar studies that were conducted using the first year~\cite{Hooper:2018wyk} and the first three years~\cite{Smith:2020oac} of IceCube data.
%
As in those previous studies, we find no evidence of neutrino emission from gamma-ray-bright AGN. We do, however, strengthen the limits on the neutrino flux from gamma-ray-bright AGN by utilizing a larger IceCube dataset and by incorporating a larger AGN catalog, as provided by the Fermi Collaboration~\cite{4lac_dr3}. These results rule out blazars as a major contributor to the diffuse high-energy neutrino flux. In contrast, it remains possible that non-blazar AGN could generate much of this diffuse flux.

In addition to gamma-ray source catalogs, we have performed a search for neutrino emission from X-ray-bright AGN, as found in the 70-month Swift Burst Alert Telescope (Swift-BAT) hard X-ray survey~\cite{swift_bat_70}; for earlier work, see Ref.~\cite{IceCube:2024dou,IceCube:2024ayt}.\footnote{Our study of the neutrino emission from X-ray emitting AGN has much in common with that of Ref.~\cite{IceCube:2024ayt}. Whereas that work employed a stacked, frequentist search for neutrino events from a predefined source list, we adopt a model-based statistical framework aimed at inferring properties of the underlying source population. In particular, our approach explicitly allows for source-to-source variation in the neutrino emission from individual AGN.} We find evidence of neutrino emission from this source population with a significance of 4.2$\sigma$, and conclude that non-blazar, X-ray-bright AGN produce a significant fraction, between 11.2\% and 100\% at $2\sigma$ confidence, of the diffuse flux observed by IceCube. Much of this statistical significance is driven by NGC 1068, which is included in this population. Our upper limits on this contribution, however, do not appreciably change if NGC 1068 is removed from our sample.

The remainder of this paper is structured as follows. In Sec.~\ref{catalogs_hypos}, we describe the AGN populations that we consider and outline the various hypotheses that we test in our study. In Sec.~\ref{data_likelihood}, we describe our analysis and statistical approach. In Sec.~\ref{results}, we present our main results for X-ray-bright AGN, gamma-ray-bright blazars, and gamma-ray-bright
 non-blazar AGN. We discuss our results and their implications for high-energy astrophysics in Sec.~\ref{discussion}, and summarize our main conclusions in Sec.~\ref{conclusions}. This paper includes two appendices which describe additional details of our analysis.

\section{Source Catalogs and Hypotheses}
\label{catalogs_hypos}

In this study, we consider several populations of AGN, as found in the 70-month Swift-BAT hard X-ray survey~\cite{swift_bat_70} and in the Fermi 4LAC-DR3 catalog~\cite{4lac_dr3}. For each source population considered, we test the following hypotheses for the predicted neutrino emission:
\begin{enumerate}
    \item{X-ray Scaling: The intensity of the neutrino flux from a given source is taken to be proportional to its intrinsic X-ray energy flux. We consider the soft and hard X-ray bands independently. The intrinsic X-ray fluxes are obtained from Ref.~\cite{obs_to_intr_x}, which corrects for the effects of absorption.}
    \item{Gamma-Ray Scaling: The intensity of the neutrino flux from a given source is taken to be proportional to its gamma-ray flux. This hypothesis corresponds to scenarios in which the gamma-ray emission from these sources is largely hadronic in origin and is not strongly obscured.}
    \item{Geometric Scaling: The intensity of the neutrino flux from a given source is taken to be proportional to $1/D_L^2$, where $D_L$ is the luminosity distance of the source. This hypothesis treats all sources as equally likely to be luminous neutrino emitters.}
\end{enumerate}

The source catalogs we utilize in this study are not complete, in that there are many AGN that they do not contain. For each source sample and hypothesis we consider, we compute a completeness factor which allows us to translate the constraint on a given source sample to a constraint on the entire source class in question. These completeness factors are described in Appendix~\ref{completeness_factors}.

In testing the X-ray scaling hypothesis, as described above, we consider the 732 non-blazar AGN found in the 70-month Swift-BAT hard X-ray survey. The intrinsic X-ray luminosities, in both the soft and hard X-ray bands, were evaluated in Ref.~\cite{obs_to_intr_x} for each of the sources in the 70-month survey. While more recent X-ray data releases are available, their intrinsic fluxes have not been evaluated. All of the sources in this catalog have reported redshifts, which we use in testing the geometric scaling hypothesis. The exceptionally bright neutrino source, NGC 1068, is included in this X-ray catalog. We present results both including and excluding it from our sample.

For the gamma-ray scaling hypothesis, we consider seperately the 3339 blazars and 64 non-blazar AGN found within the Fermi 4LAC-DR3 catalog~\cite{4lac_dr3,4lac_original}. We also perform tests on the non-variable subset of this catalog, which consists of 1536 blazars and 40 non-blazar AGN.\footnote{Following the Fermi Collaboration, we classify a source as variable if it has a reported variability index greater than 18.48, corresponding to evidence of variability at the 99\% confidence level~\cite{4lac_original}.} In each of these cases, only a subset of the gamma-ray-bright AGN have reported redshifts; in testing the geometric scaling hypothesis, we limit ourselves to this smaller sample; see Table~\ref{table_4lac_nsrcs}.

\begin{table}[t]
\centering
\begin{tabular}{l|c|c}
\hline
Source Sample & All & With Redshift \\
\hline
All Blazars & 3339 & 1749 \\
Non-Variable Blazars & 1536 & 653 \\
All Non-Blazar AGN & 64 & 57 \\
Non-Variable, Non-Blazar AGN & 40 & 34 \\
\hline
\end{tabular}
\caption{The number of AGN in each of our gamma-ray source samples, as contained in Fermi's 4LAC-DR3 catalog. \label{table_4lac_nsrcs}}
\end{table}

\section{Data and Likelihood Estimation}
\label{data_likelihood}

In this study, we utilize the most recent publicly available IceCube dataset, consisting of muon-track neutrino candidate events accumulated over 10 years of operation between 2008 and 2018~\cite{icecube_data}. For each event, the energy and direction of the muon, as well as the uncertainty of its arrival direction, are reported. For each of the detector configurations between 2008 and 2018, the data release includes its energy- and direction-dependent neutrino effective area. To make use of the reported energies of the muon tracks, we convert the neutrino effective area into a muon effective area, as described in Appendix~\ref{muon_area}.

We consider muon-tracks with energies between $10^2$ to $10^8$ GeV. For each energy bin, $j$, we construct the following likelihood for the total number of events in the detector, $n_{s,j}$, from a given direction in the sky, $\vec{x}_s$: 
\begin{equation}
    \mathcal{L}(n_{s,j}, \vec{x}_s) = \prod_i^{N_j} \left[ \frac{n_{s,j}}{N_j} S_i(|\vec{x}_s - \vec{x}_i|) + \left(1 - \frac{n_{s,j}}{N_j}\right) B_i(\sin \delta_i) \right],
\end{equation}
where the sum is performed over all of the events in the energy bin, $N_j$ is the total number of events in that bin, and $\vec{x}_i$ and $\delta_i$ are the direction and declination of event $i$, respectively. $S_i$ and $B_i$ are the signal and background probability distribution functions (PDFs), which are defined as follows~\cite{Smith:2020oac}:
\begin{align}
    S_i(|\vec{x}_s - \vec{x}_i|) &= \frac{1}{2\pi \sigma_i^2} \, \exp\left(-\frac{|\vec{x}_s - \vec{x}_i|^2}{2\sigma_i^2}\right), \\
B_i(\sin \delta_i) &= \frac{\mathcal{P}_B(\sin \delta_i)}{2\pi},
\end{align}
where $\sigma_i$ is the angular resolution of event $i$, and 
\begin{equation}
    \mathcal{P}_B(\sin \delta_i) =\frac{N_j|^{\delta_i + 3}_{\delta_i-3}}{N_j} \, \frac{1}{\cos (\delta-3^\circ) - \cos (\delta+3^\circ)}.
\end{equation}
In this expression, $N_j|^{\delta_i + 3}_{\delta_i-3}$ is the total number of events in the $j$\textsuperscript{th} energy bin between declinations of $\delta_i-3^\circ$ and $\delta_i+3^\circ$. This background PDF stems from the fact that IceCube's backgrounds are dominated by atmospheric neutrinos and muons, which are a function of declination and are independent of right ascension. We use the measured flux within a $\pm 3^{\circ}$ band of declination to empirically determine this background at each point on the sky.

The likelihood that a given source is responsible for $n_s$ total events is given by the product of the likelihoods for each energy bin,
\begin{equation}
    \mathcal{L}(n_s, \vec{x}_s) = \prod_j \mathcal{L}(n_{s,j}, \vec{x}_s),
\end{equation}
where $n_s = \sum_j n_{s,j}$. Finally, the likelihood that a population of sources each produce $n_i$ events is given by
\begin{equation}
        \mathcal{L}(n_0, n_1, n_2,...) = \prod_k \mathcal{L}(n_k, \vec{x}_k).
\end{equation}

For each of the hypotheses described in Sec.~\ref{catalogs_hypos}, the neutrino flux from each source is predicted up to an overall normalization factor which is allowed to float. Realistically, one would expect there to be some degree of source-to-source variation in the neutrino flux. To allow for this, we consider a log-normal PDF for a source with a predicted flux, $F_0$, to produce an actual flux, $F$~\cite{Hooper:2018wyk,Fermi-LAT:2013sme}:
\begin{equation}
    \frac{dP}{dF} = \frac{1}{\sqrt{2\pi}\delta F} \, \exp\left(-\frac{(\ln{F/F_0})^2}{2\delta^2}\right),
\end{equation}
where the standard deviation of this distribution, $\delta$, is treated as a free parameter and should not be confused with declination. Note that since this PDF is a gaussian in log-space, the mean value of the flux is not $F_0$, but instead is given by 
\begin{equation}
    \langle F \rangle = \int_0^\infty F \frac{dP}{dF} dF = F_0 \, \exp\left(\frac{\delta^2}{2}\right).
\end{equation}
We thus define our final likelihood as
\begin{equation}
    \mathcal{L}(\vec{F}, \delta) = \prod_k \int_0^\infty \mathcal{L}(F_k^\prime) \, \frac{1}{\sqrt{2\pi} \, \delta \, F_k^\prime} \, \exp{\left(-\frac{(\ln{F_k^\prime/F_k^0})^2}{2\delta^2}\right)} \, dF_k^\prime,
\end{equation}
where $F_k^0 = F_k \exp\left(-\delta^2 /2\right)$, and $\vec{F} = F_0, F_1,...$ represents the fluxes from the members of the source population. The flux from a given source is converted into a mean number of predicted muon-track events using the muon effective area calculated in Appendix~\ref{muon_area}. 

Finally, our test statistic (TS) is defined as twice the change in the log-likelihood,

\begin{equation}
    {\rm TS}(\vec{F}, \delta) = 2\ln\left( \frac{\mathcal{L}(\vec{F}, \delta)}{\mathcal{L}(\vec{F} = 0)}\right).
    \label{mean_flux}
\end{equation}
From the TS value, we calculate the corresponding statistical significance using the appropriate number of degrees-of-freedom.

\begin{figure}[t]
\centering
\includegraphics[width=0.65\linewidth]{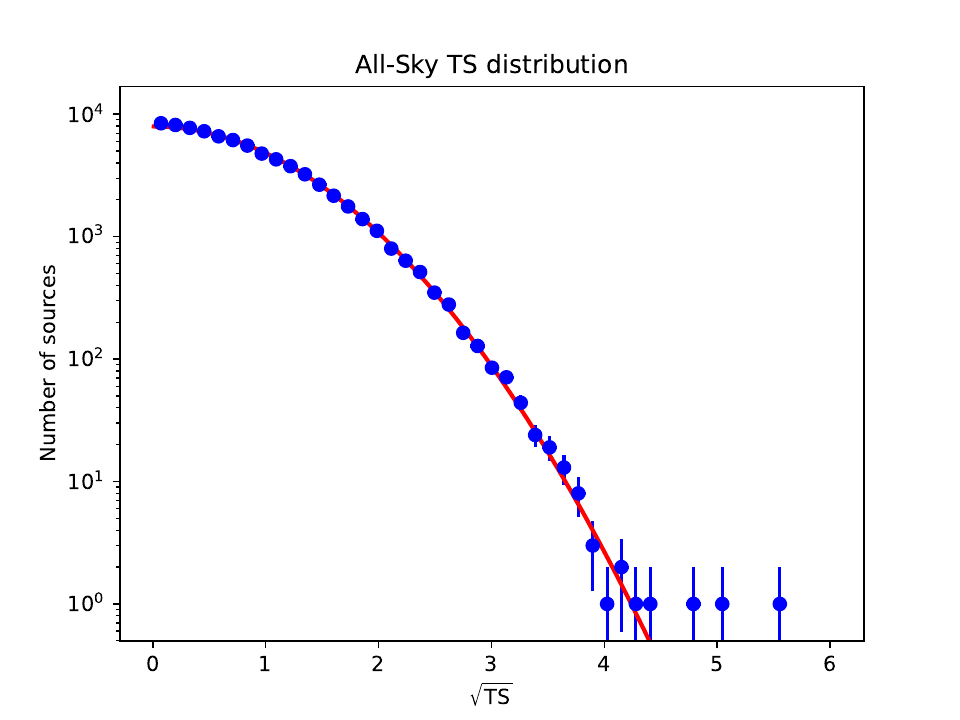}
\caption{The distribution of our test statistic (TS) at 196,608 points evenly spaced across the sky, adopting a spectral index of 2.5. This result follows a gaussian distribution and reveals no evidence of any neutrino source population.}
\label{all_sky_TS_distribution}
\end{figure}

Throughout our analysis, we take the neutrino spectrum of AGN to be described by a power-law, $dN_{\nu}/dE_{\nu} \propto E^{-\alpha}_{\nu}$, and consider two values of the spectral index, $\alpha =2.5$ and 3.0. The former value is motivated by the measured index of IceCube's diffuse neutrino flux~\cite{icecube_diffuse_recent}, while the latter value is suggested by the steeper measured spectrum of NGC 1068; see Table \ref{brightest}. We take the overall normalization of the neutrino flux from a source population to be a free parameter, and similarly allow the degree of source-to-source variation to vary between $\delta = 0$ and 2.5.\footnote{While an index of $\alpha=2.0$ could be motivated by considerations related to first-order Fermi acceleration, we find that most of the Swift-BAT sources which yield a high TS value are best fit by a softer spectral index, $\alpha \ge 2.5$.}

In Fig.~\ref{all_sky_TS_distribution}, we plot the distribution of our test statistic for 196,608 points evenly spaced on the sky, for a spectral index of 2.5. This result follows a gaussian distribution, revealing no evidence of any neutrino source population.



\section{Results}
\label{results}

In this section, we describe the main results of our analysis. 
For each case, 
we derive constraints on the total neutrino flux from the source population evaluated at $E_{\nu} =30\,{\rm TeV}$, and on the source-to-source variation in this flux, $\delta$. The fluxes shown in these figures include a completeness factor which accounts for the fraction of the source population that is not contained in the catalog; see Appendix~\ref{completeness_factors}. We also plot the TS profile as a function of each of these parameters.





\begin{figure}[t]
\centering
\begin{minipage}{0.5\textwidth}
\centering
\includegraphics[width=1.0\linewidth]{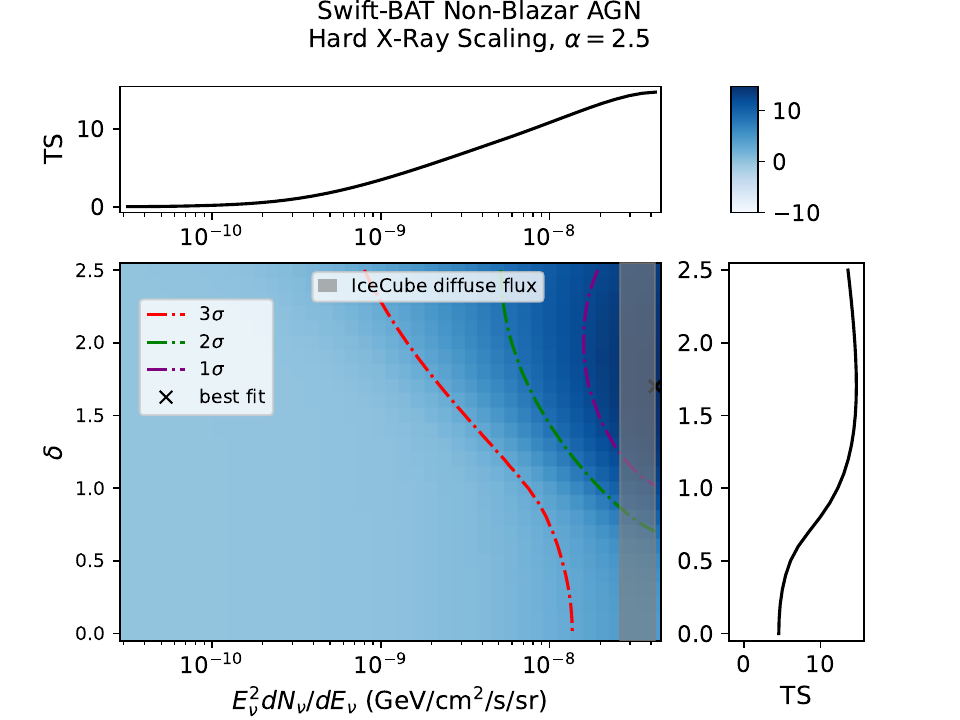}
\end{minipage}\hfill
\begin{minipage}{0.5\textwidth}
\centering
\includegraphics[width=1.0\linewidth]{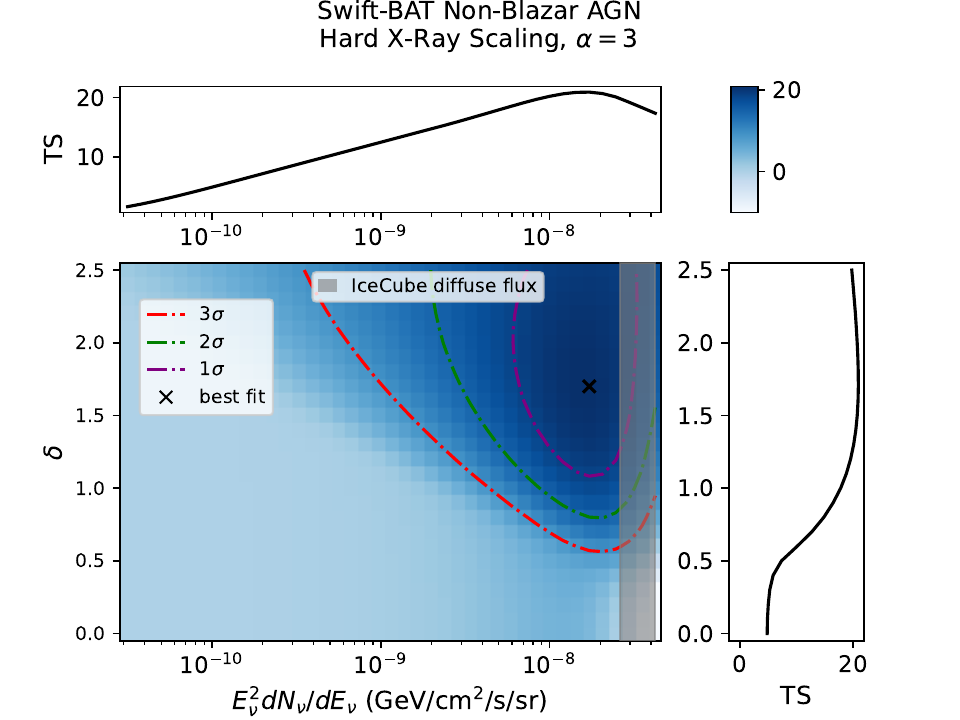}
\end{minipage}

\vspace{0.6cm}

\begin{minipage}{0.5\textwidth}
\centering
\includegraphics[width=1.0\linewidth]{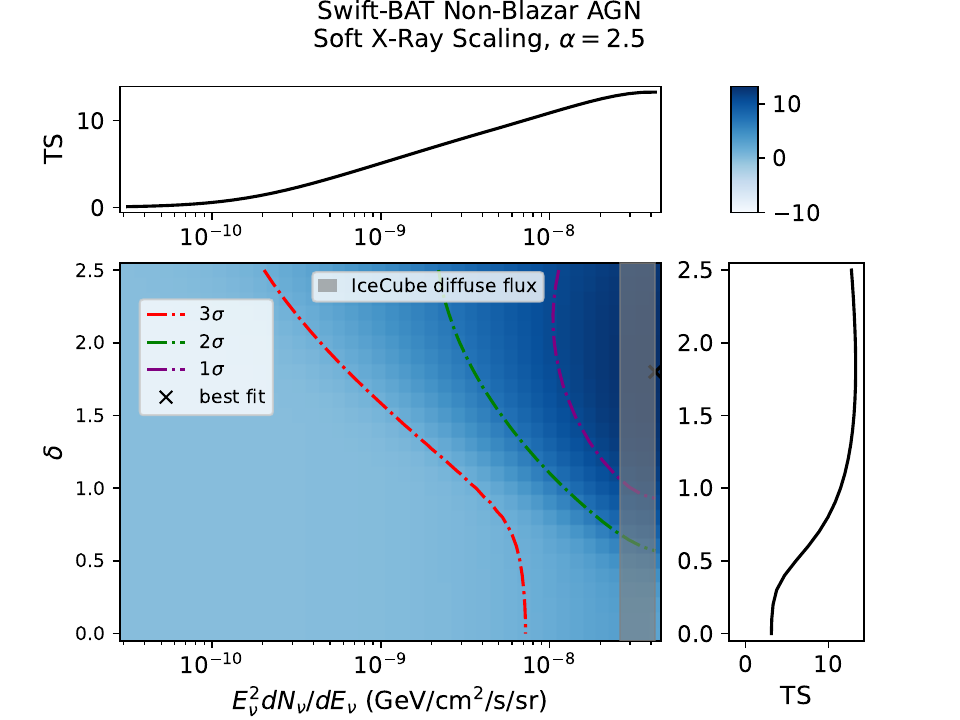}
\end{minipage}\hfill
\begin{minipage}{0.5\textwidth}
\centering
\includegraphics[width=1.0\linewidth]{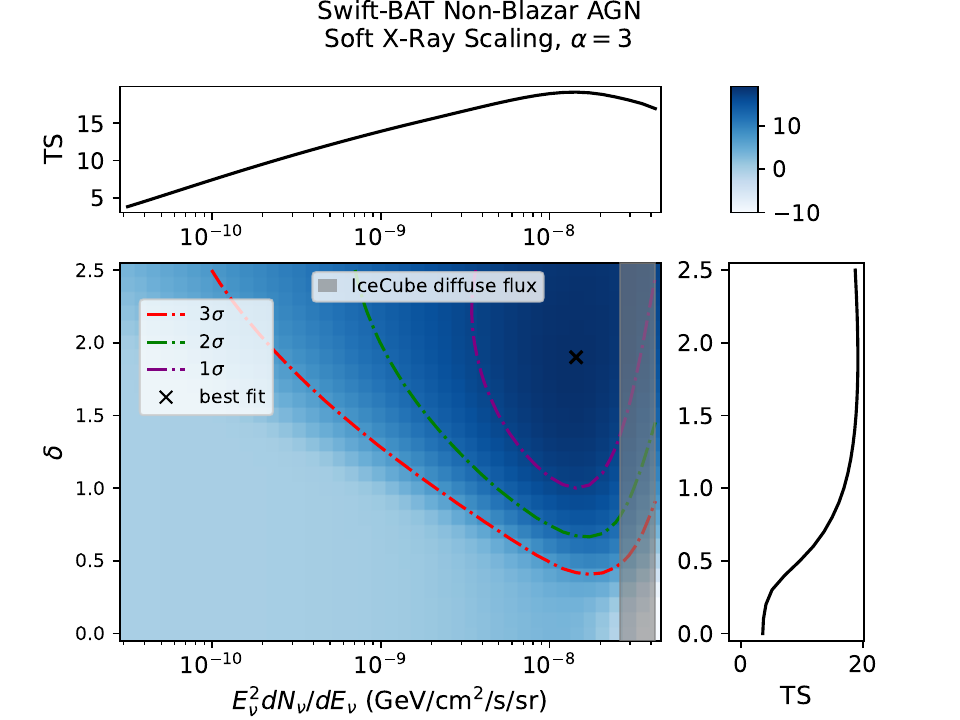}
\end{minipage}

\vspace{0.6cm}

\begin{minipage}{0.5\textwidth}
\centering
\includegraphics[width=1.0\linewidth]{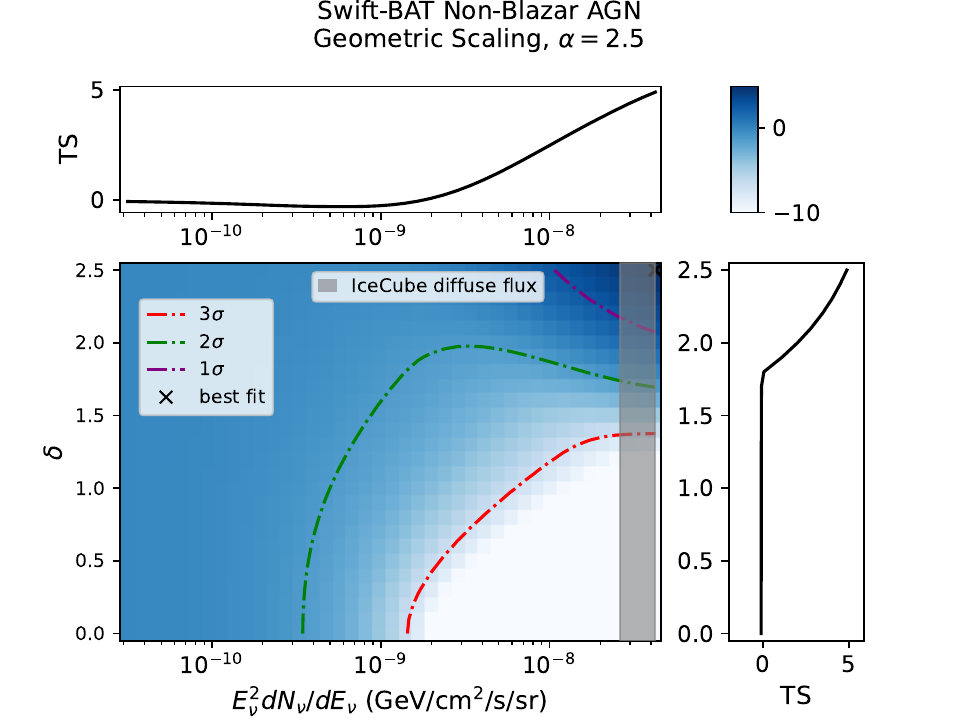}
\end{minipage}\hfill
\begin{minipage}{0.5\textwidth}
\centering
\includegraphics[width=1.0\linewidth]{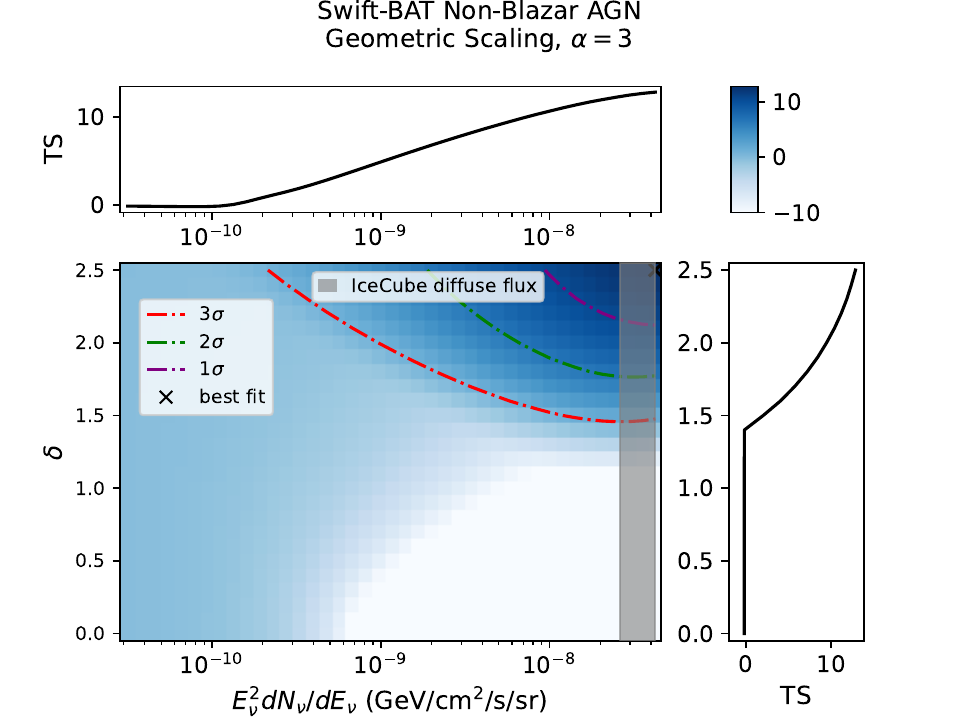}
\end{minipage}
\caption{Constraints on the total neutrino flux from X-ray-bright, non-blazar AGN (per flavor and evaluated at $E_{\nu} =30\,{\rm TeV}$), and on the source-to-source variation in this flux, $\delta$.  We also plot the test statistic (TS) profile as a function of each of these parameters. These fluxes include a completeness factor that accounts for the fraction of the source population that is not contained in the 70-month Swift-BAT catalog (see Table~\ref{completeness_factors_all}). In the top and middle rows, we show the results for the hard and soft X-ray scaling hypotheses, respectively. In the bottom row, we show the results for the geometric scaling hypothesis. In the left and right frames, we adopt a neutrino spectral index of $\alpha=2.5$ or 3.0. The vertical gray band denotes the total diffuse flux measured by IceCube~\cite{icecube_diffuse_recent}.}
\label{swift_full}
\end{figure}

\subsection{X-Ray-Bright, Non-Blazar AGN}

We begin by considering the 732 non-blazar AGN contained in the catalog of the 70-month Swift-BAT hard X-ray survey. Our main results for this case are shown in Fig.~\ref{swift_full}. For these sources, we find a statistically significant preference for neutrino emission, corresponding to $4.18 \sigma$ ($3.98\sigma$) for the case of hard (soft) X-ray scaling and $\alpha=3.0$; see Table~\ref{table_swift_full}. For the geometric scaling hypothesis, the statistical significance of this emission drops to $3.15\sigma$. For a harder spectral index of $\alpha=2.5$, we find less evidence of neutrino emission. Across all of the cases considered, our analysis finds that this class of sources should produce between 5$\%$ and 100$\%$ of the total diffuse flux observed by IceCube at the 2$\sigma$ confidence level. These results are broadly consistent with those presented in Ref.~\cite{IceCube:2024ayt}.

\begin{figure}[t]
\centering

\begin{minipage}{0.5\textwidth}
\centering
\includegraphics[width=1.0\linewidth]{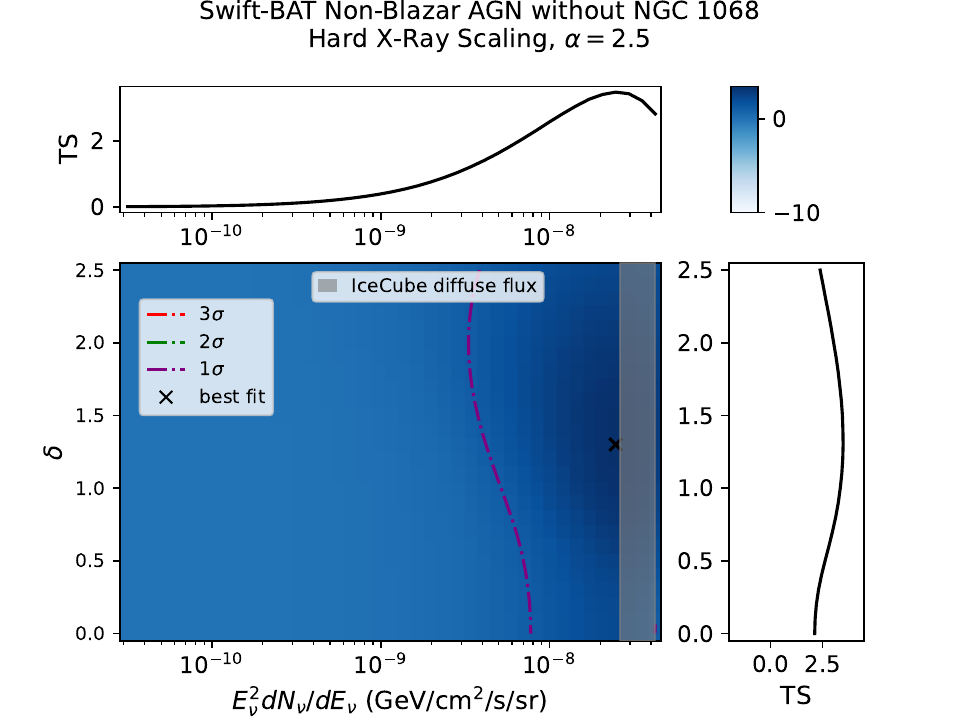}
\end{minipage}\hfill
\begin{minipage}{0.5\textwidth}
\centering
\includegraphics[width=1.0\linewidth]{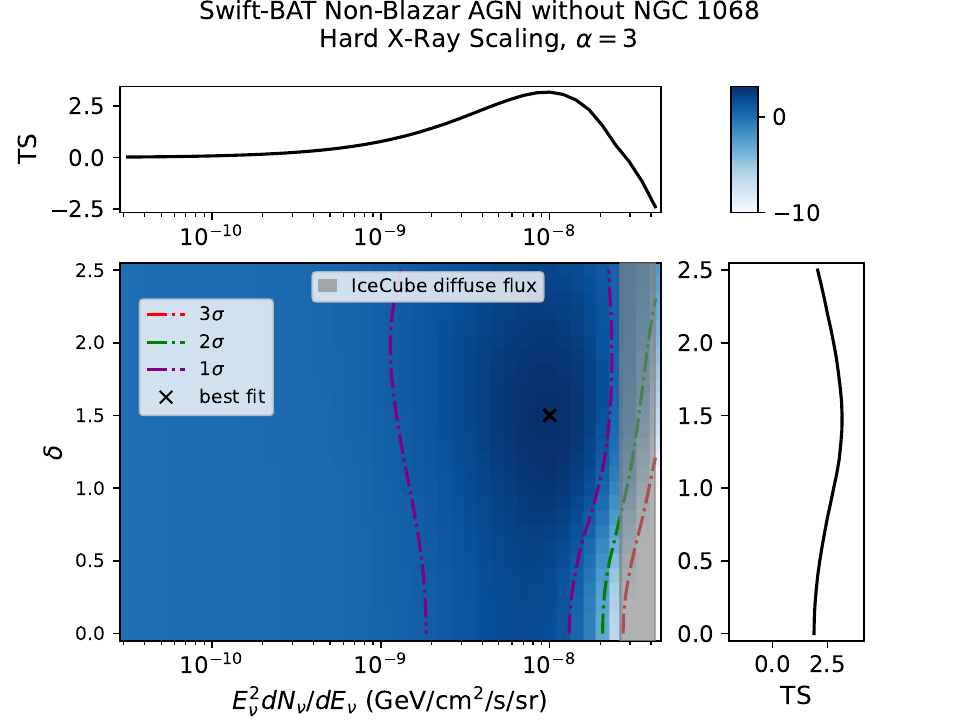}
\end{minipage}

\vspace{0.6cm}

\begin{minipage}{0.5\textwidth}
\centering
\includegraphics[width=1.0\linewidth]{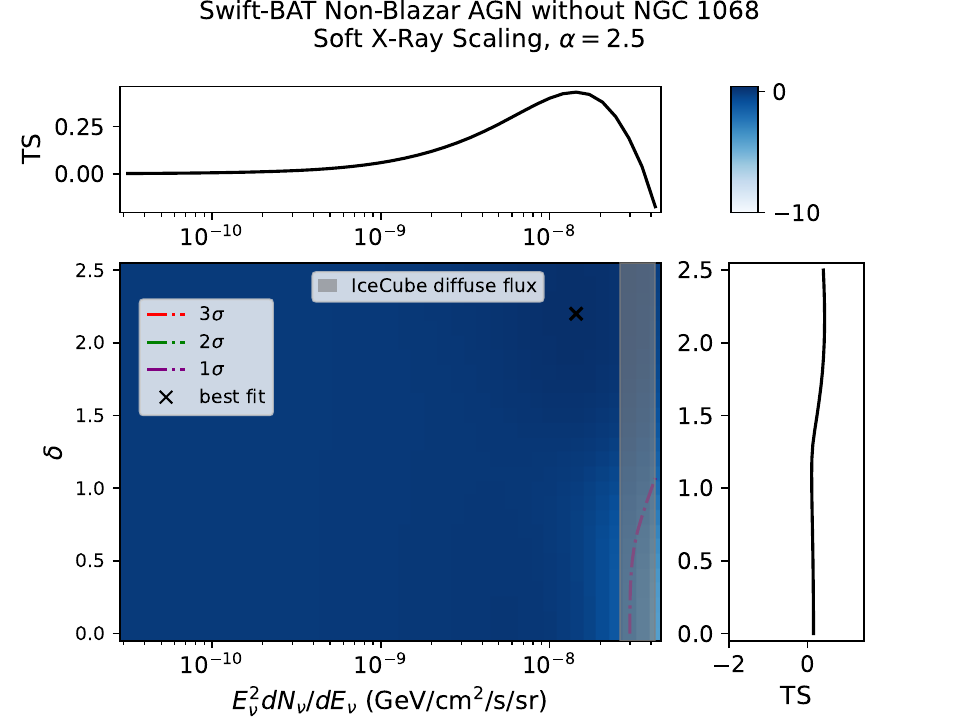}
\end{minipage}\hfill
\begin{minipage}{0.5\textwidth}
\centering
\includegraphics[width=1.0\linewidth]{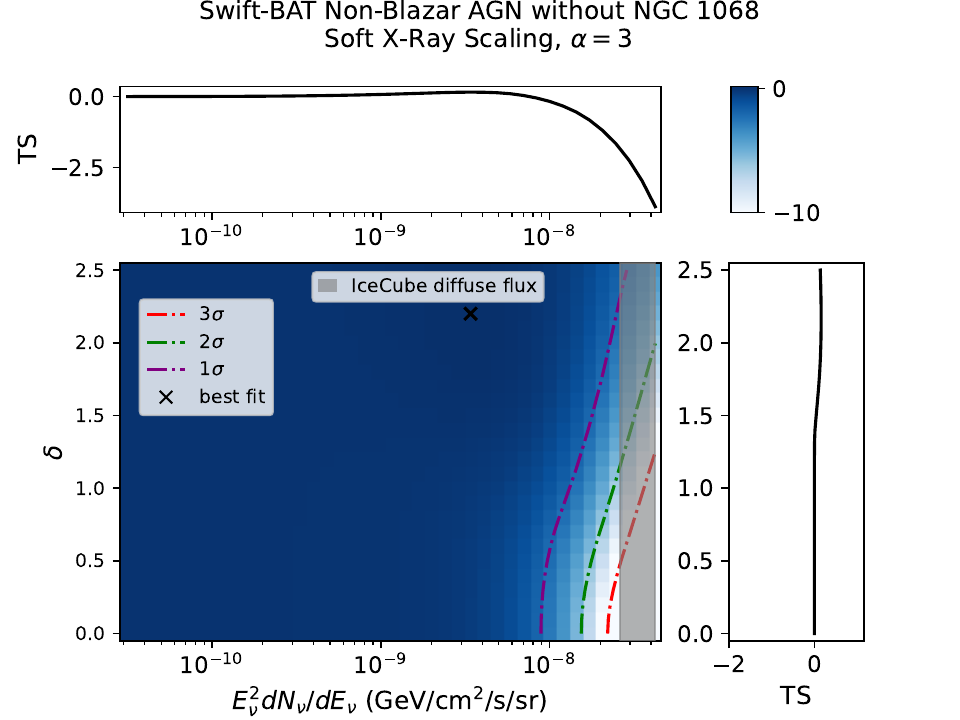}
\end{minipage}

\vspace{0.6cm}

\begin{minipage}{0.5\textwidth}
\centering
\includegraphics[width=1.0\linewidth]{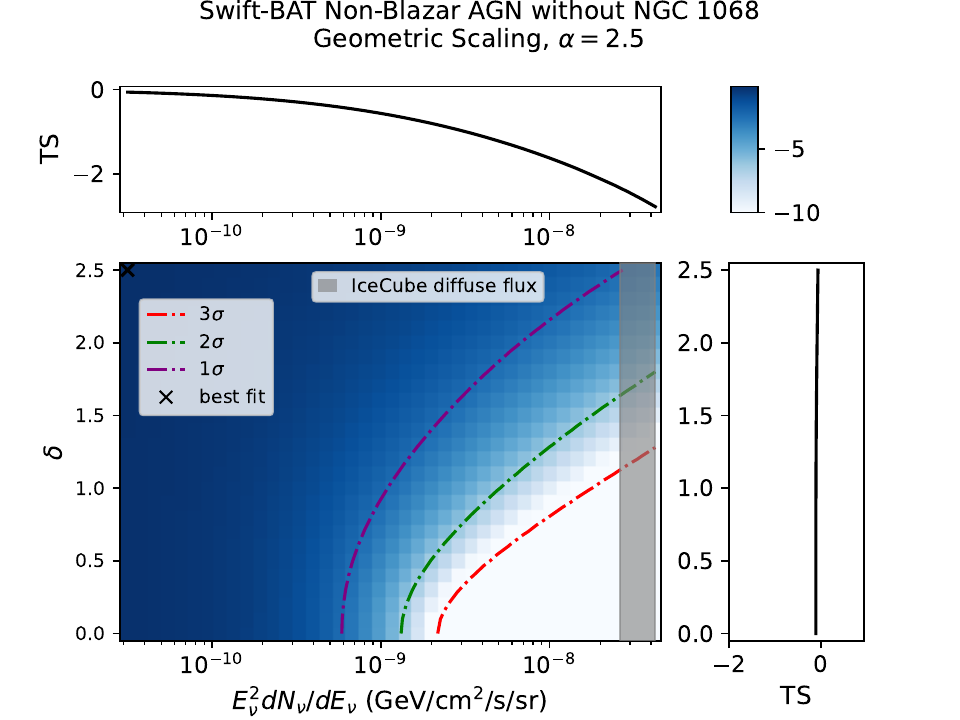}
\end{minipage}\hfill
\begin{minipage}{0.5\textwidth}
\centering
\includegraphics[width=1.0\linewidth]{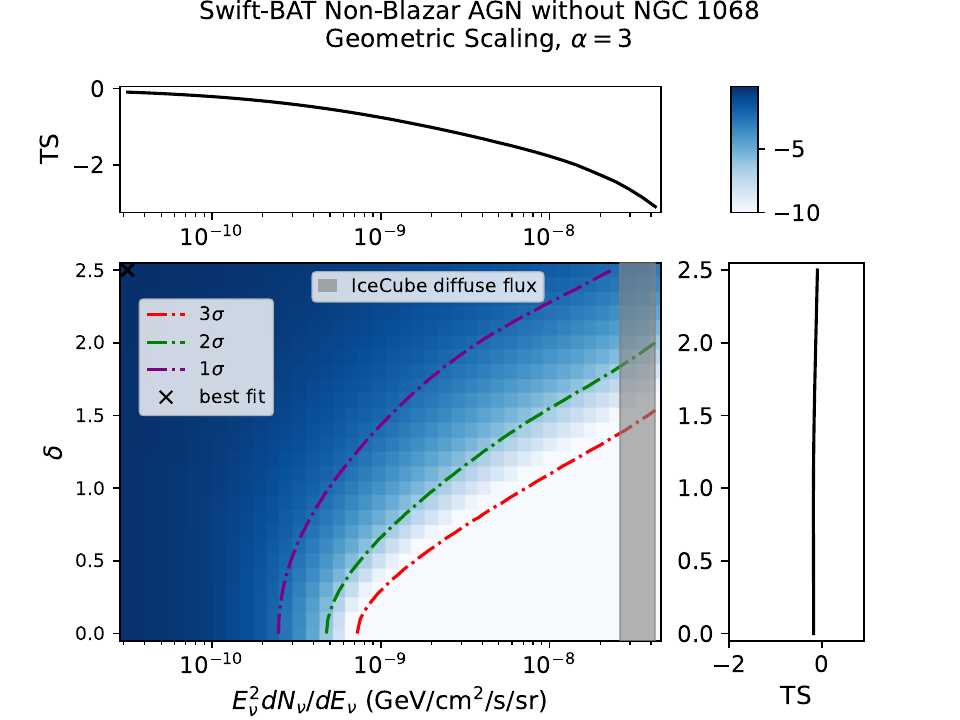}
\end{minipage}

\caption{As in Fig.~\ref{swift_full}, but excluding NGC 1068.\\}
\label{swift_without_ngc1068}

\end{figure}

Much of the statistical significance for neutrino emission from this source population is the result of a single source, NGC 1068. If we exclude this source from our analysis, the total significance of neutrino emission from this AGN population drops from $4.18\sigma$ to $1.35\sigma$. This does not, however, significantly impact our upper limits on the total neutrino flux from this source population; see Fig.~\ref{swift_without_ngc1068} and Table~\ref{table_swift_without_ngc1068}. 

We note that the IceCube Collaboration carried out a similar analysis~\cite{icecube_hard} to that presented here, but which did not allow for source-to-source variation in the expected neutrino flux. Our results with $\delta=0$ are consistent with those of that study.



\clearpage

\begin{table}[H]
\centering
\textbf{X-Ray-Bright, Non-Blazar AGN (Swift-BAT)} \\
\vspace{0.3cm}
\begin{tabular}{c|c|c}
Hypothesis & Significance~($\sigma$) & Contribution to $F^\text{diffuse}_{\nu} \, (2\sigma)$  \\
\hline
Hard X-Ray Scaling, $\alpha=2.5$ & 3.41 & $F^{\rm AGN}_{\nu}> 28.9\% $\\
\hline
Soft X-Ray Scaling, $\alpha=2.5$ & 3.21 & $F^{\rm AGN}_{\nu}> 15.7\% $  \\ 
\hline
Geometric Scaling, $\alpha=2.5$ & 1.71 &  $F^{\rm AGN}_{\nu} > 11.8\% $  \\
\hline
Hard X-Ray Scaling, $\alpha=3.0$ & 4.18 & $F^{\rm AGN}_{\nu}> 11.2\%$ \\
\hline
Soft X-Ray Scaling, $\alpha=3.0$ & 3.98 & 
$F^{\rm AGN}_{\nu}>5.0\% $ \\ 
\hline
Geometric Scaling, $\alpha = 3.0$ & 3.15 & 
$F^{\rm AGN}_{\nu}>13.1\%$\\
\end{tabular}
\caption{A summary of our results for X-ray-bright, non-blazar AGN. The constraints on the contribution to IceCube's diffuse neutrino flux are presented in terms of the flux from the entire source class, obtained after applying an appropriate completeness factor (see Table~\ref{completeness_factors_all}). \label{table_swift_full}}
\end{table}

\vspace{2cm}

\begin{table}[H]
\centering
\textbf{X-Ray-Bright, Non-Blazar AGN (Swift-BAT), Excluding NGC 1068} \\
\vspace{0.3cm}
\begin{tabular}{c|c|c}
Hypothesis & Significance~($\sigma$) & Contribution to $F^\text{diffuse}_{\nu} \, (2\sigma)$ \\
\hline
Hard X-Ray Scaling, $\alpha=2.5$ & 1.35 & No Constraint  \\
\hline
Soft X-Ray Scaling, $\alpha=2.5$ & 0.25 &  No Constraint  \\ 
\hline
Geometric Scaling, $\alpha=2.5$ & 0.0 & No Constraint  \\
\hline
Hard X-Ray Scaling, $\alpha=3.0$ & 1.27 & 
$F^{\rm AGN}_{\nu} < 99.1\%$ \\
\hline
Soft X-Ray Scaling, $\alpha = 3.0$ & 0.10 & No Constraint \\ 
\hline
Geometric Scaling, $\alpha = 3.0$ & 0.0 & No Constraint\\
\end{tabular}
\caption{As in Table~\ref{table_swift_full}, but excluding NGC 1068.}
\label{table_swift_without_ngc1068}
\end{table}

\vspace{1cm}

\subsection{Gamma-Ray-Bright Blazars}
\begin{figure}[h]
\centering

\begin{minipage}{0.5\textwidth}
\centering
\includegraphics[width=1.0\linewidth]{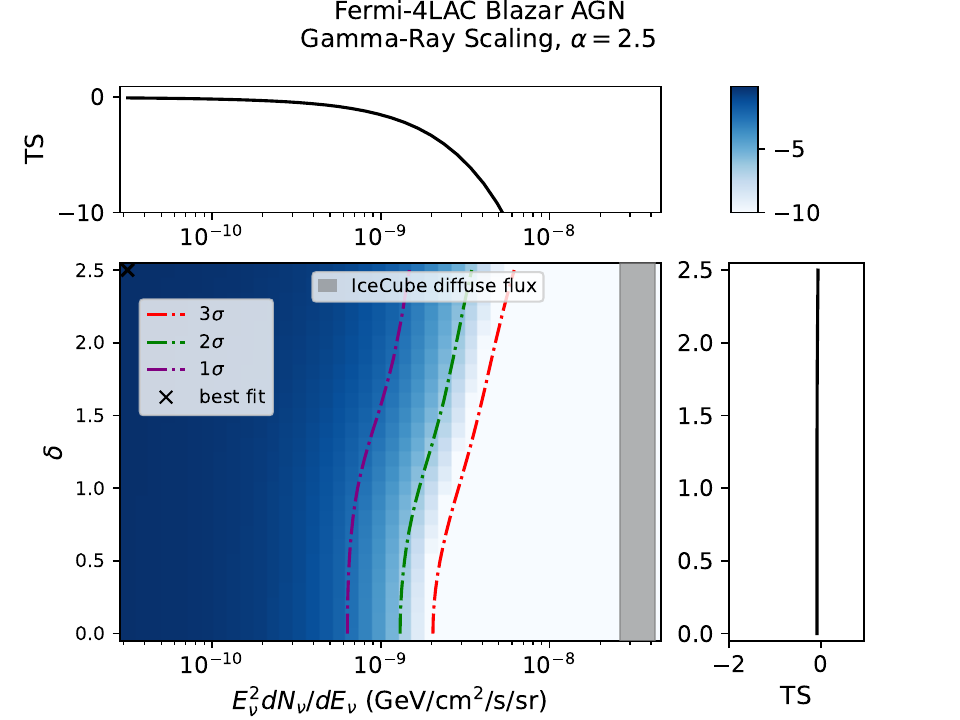}
\end{minipage}\hfill
\begin{minipage}{0.5\textwidth}
\centering
\includegraphics[width=1.0\linewidth]{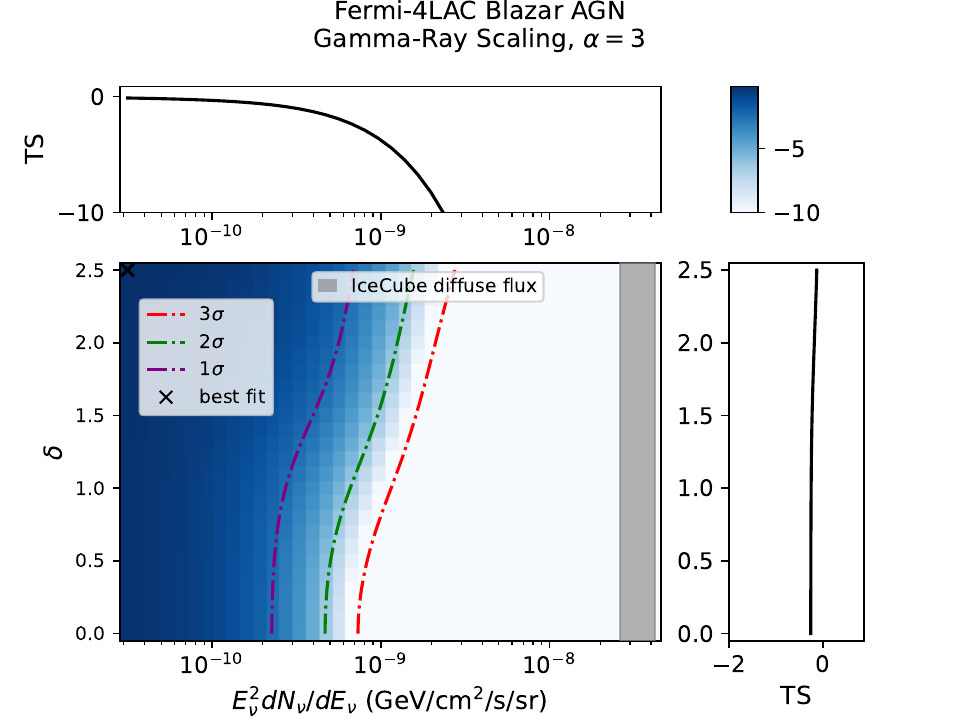}
\end{minipage}

\vspace{0.6cm}

\begin{minipage}{0.5\textwidth}
\centering
\includegraphics[width=1.0\linewidth]{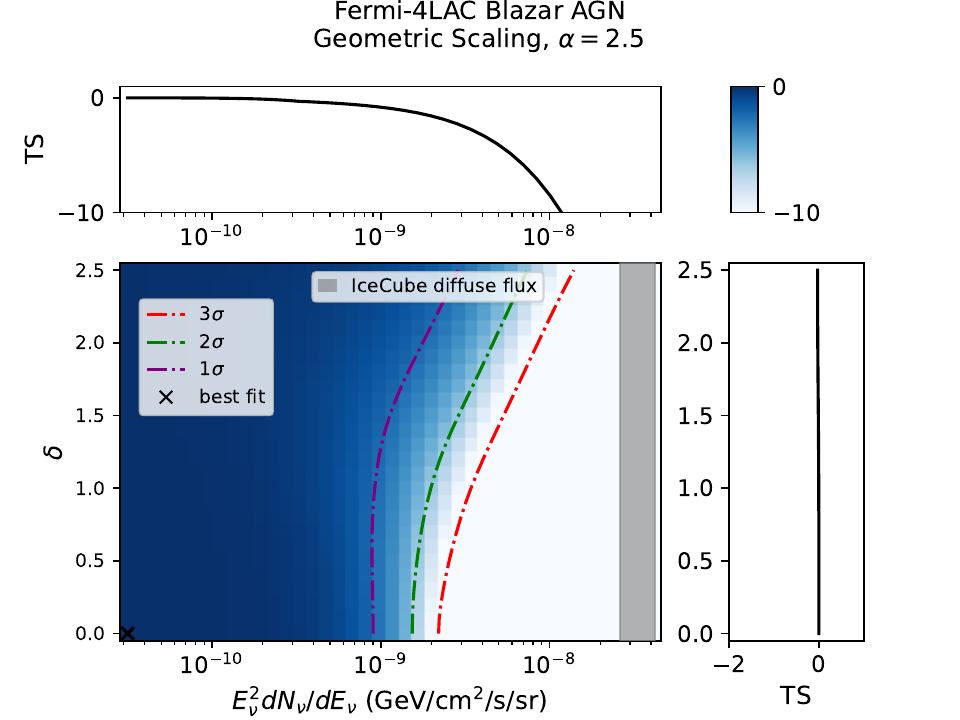}
\end{minipage}\hfill
\begin{minipage}{0.5\textwidth}
\centering
\includegraphics[width=1.0\linewidth]{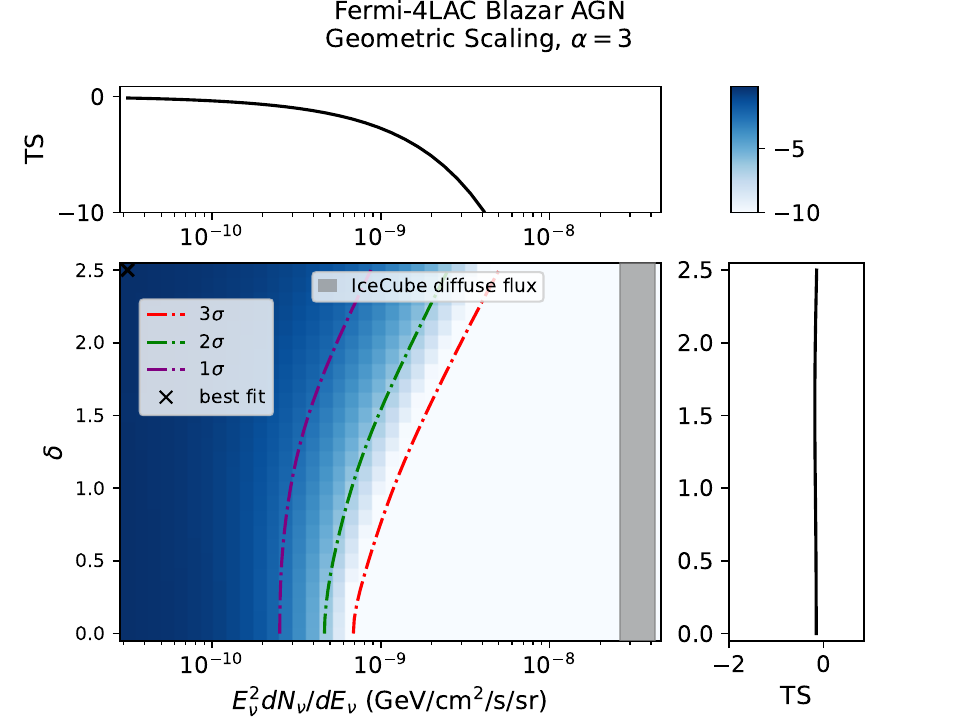}
\end{minipage}

\caption{Constraints on the total neutrino flux from gamma-ray-bright, blazar AGN (per flavor and evaluated at $E_{\nu} =30\,{\rm TeV}$), and on the source-to-source variation in this flux, $\delta$. These fluxes include a modest completeness factor that accounts for the fraction of the source population that is not contained in the Fermi 4LAC-DR3 catalog (see Table~\ref{completeness_factors_all}). We also plot the test statistic (TS) profile as a function of each of these parameters. In the top rows, we show the results for the gamma-ray scaling hypotheses. In the bottom row, we show the results for the geometric scaling hypothesis. In the left and right frames, we have adopted a neutrino spectral index of $\alpha=2.5$ and 3.0. The vertical gray band denotes the total diffuse flux measured by IceCube~\cite{icecube_diffuse_recent}.}
\label{4LAC_all_blazars_fig}
\end{figure}

\begin{figure}[H]
\centering

\begin{minipage}{0.5\textwidth}
\centering
\includegraphics[width=1.1\linewidth]{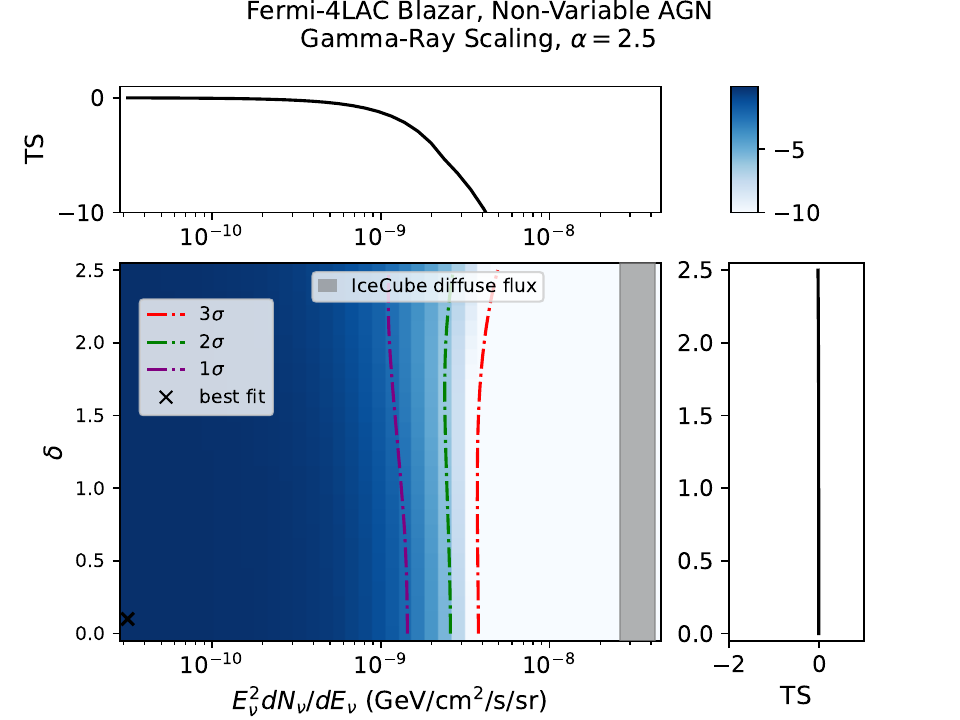}
\end{minipage}\hfill
\begin{minipage}{0.5\textwidth}
\centering
\includegraphics[width=1.1\linewidth]{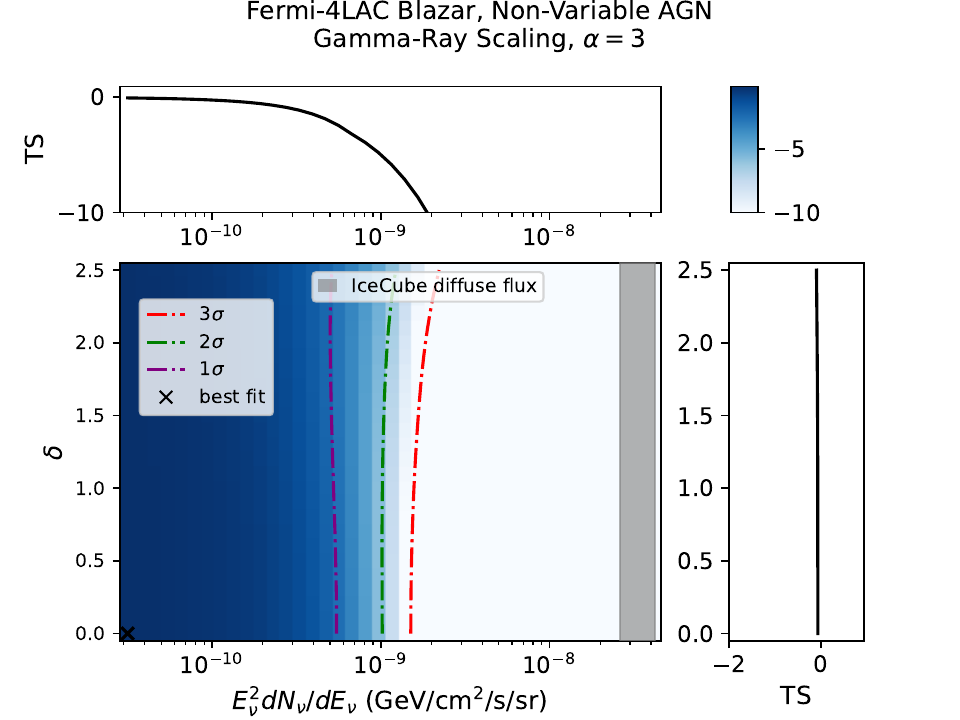}
\end{minipage}

\vspace{0.7cm}

\begin{minipage}{0.5\textwidth}
\centering
\includegraphics[width=1.1\linewidth]{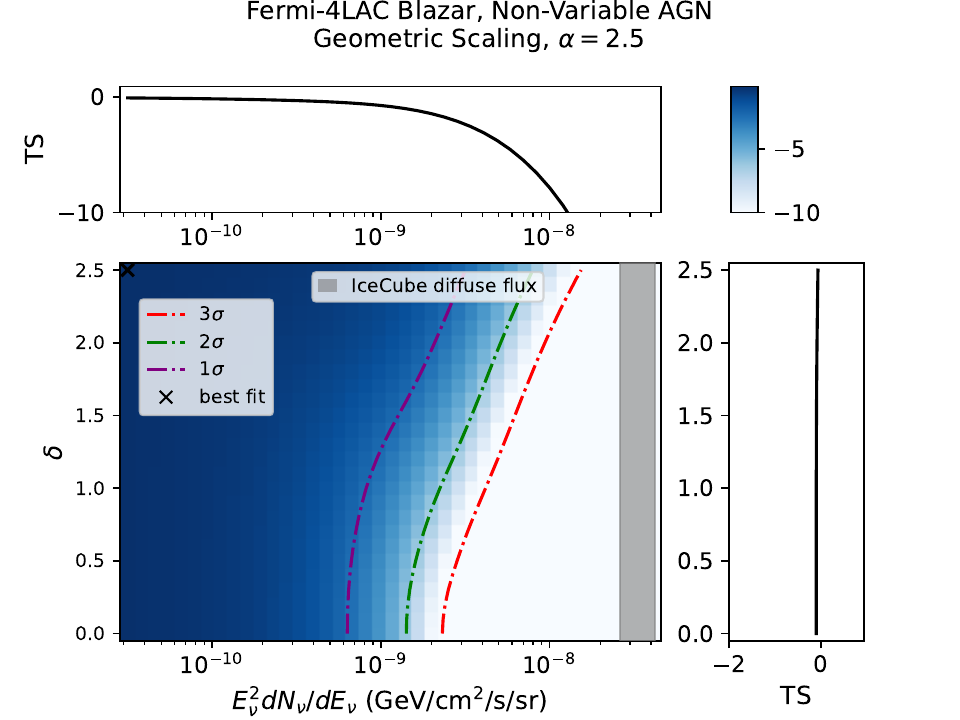}
\end{minipage}\hfill
\begin{minipage}{0.5\textwidth}
\centering
\includegraphics[width=1.1\linewidth]{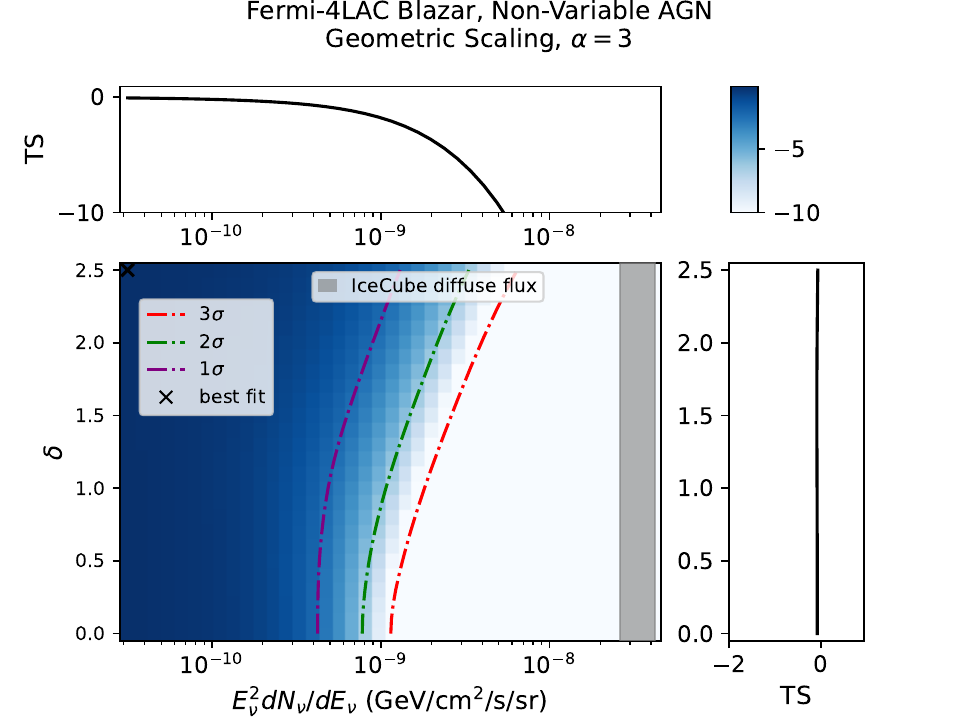}
\end{minipage}

\caption{As in Fig.~\ref{4LAC_all_blazars_fig}, but restricting our sample to non-variable gamma-ray-bright AGN.}
\label{4LAC_nonvar_blazars_fig}
\end{figure}

\vspace{0.5cm}
Next, we consider the 3339 blazars contained in the Fermi 4LAC-DR3 catalog (of which 1536 are non-variable). Our results in this case are shown in Fig.~\ref{4LAC_all_blazars_fig} and Fig.~\ref{4LAC_nonvar_blazars_fig} for the total and non-variable populations, respectively. In each of the scenarios we have tested, we identify no evidence of neutrino emission, and place stringent constraints on the contribution from these sources to the diffuse high-energy neutrino flux. More specifically, we conclude that no more than 14.5\% (15.6\%) of the diffuse flux measured by IceCube can be produced by all (all non-variable) gamma-ray-bright blazars. These constraints are summarized in Tables \ref{4LAC_all_blazars_table} and~\ref{4LAC_nonvar_blazars_table}.

\subsection{Gamma-Ray-Bright, Non-Blazar AGN}

\begin{figure}[H]
\centering

\begin{minipage}{0.5\textwidth}
\centering
\includegraphics[width=1.0\linewidth]{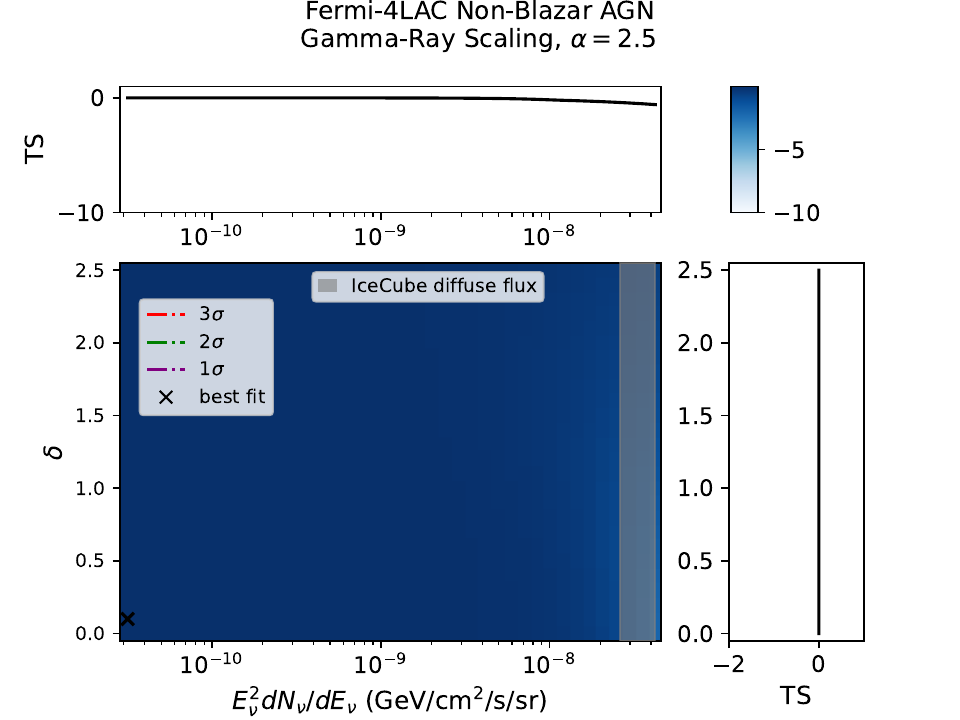}
\end{minipage}\hfill
\begin{minipage}{0.5\textwidth}
\centering
\includegraphics[width=1.0\linewidth]{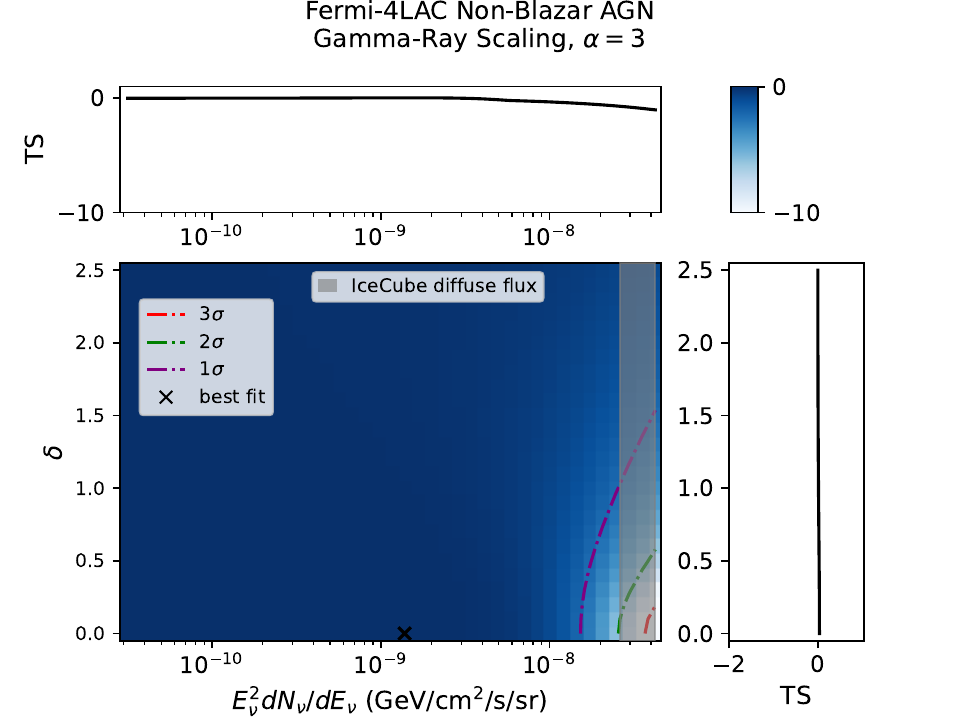}
\end{minipage}

\vspace{0.6cm}

\begin{minipage}{0.5\textwidth}
\centering
\includegraphics[width=1.0\linewidth]{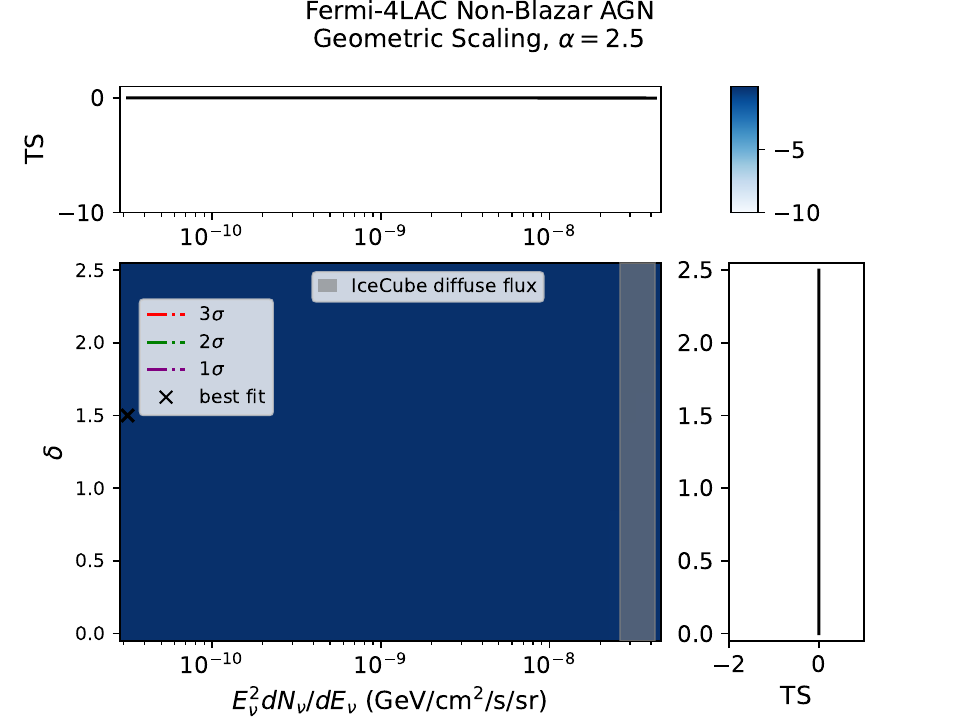}
\end{minipage}\hfill
\begin{minipage}{0.5\textwidth}
\centering
\includegraphics[width=1.0\linewidth]{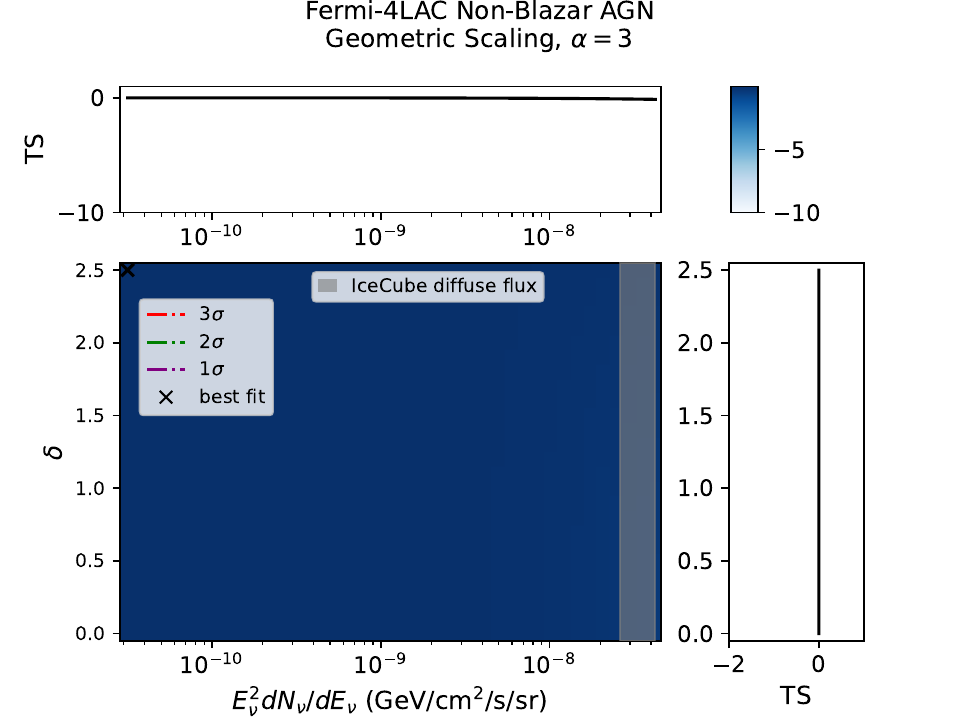}
\end{minipage}

\caption{Constraints on the total neutrino flux from gamma-ray-bright, non-blazar AGN (per flavor and evaluated at $E_{\nu} =30\,{\rm TeV}$), and on the source-to-source variation in this flux, $\delta$. These fluxes include a modest completeness factor that accounts for the fraction of the source population that is not contained in the Fermi 4LAC-DR3 catalog (see Table~\ref{completeness_factors_all}). We also plot the test statistic (TS) profile as a function of each of these parameters. In the top rows, we show the results for the gamma-ray scaling hypotheses. In the bottom row, we show the results for the geometric scaling hypothesis. In the left and right frames, we have adopted a neutrino spectral index of $\alpha=2.5$ and 3.0. The vertical gray band denotes the total diffuse flux measured by IceCube~\cite{icecube_diffuse_recent}.}
\label{4LAC_all_nonblazars_fig}
\end{figure}
\begin{figure}[H]
\centering

\begin{minipage}{0.5\textwidth}
\centering
\includegraphics[width=1.0\linewidth]{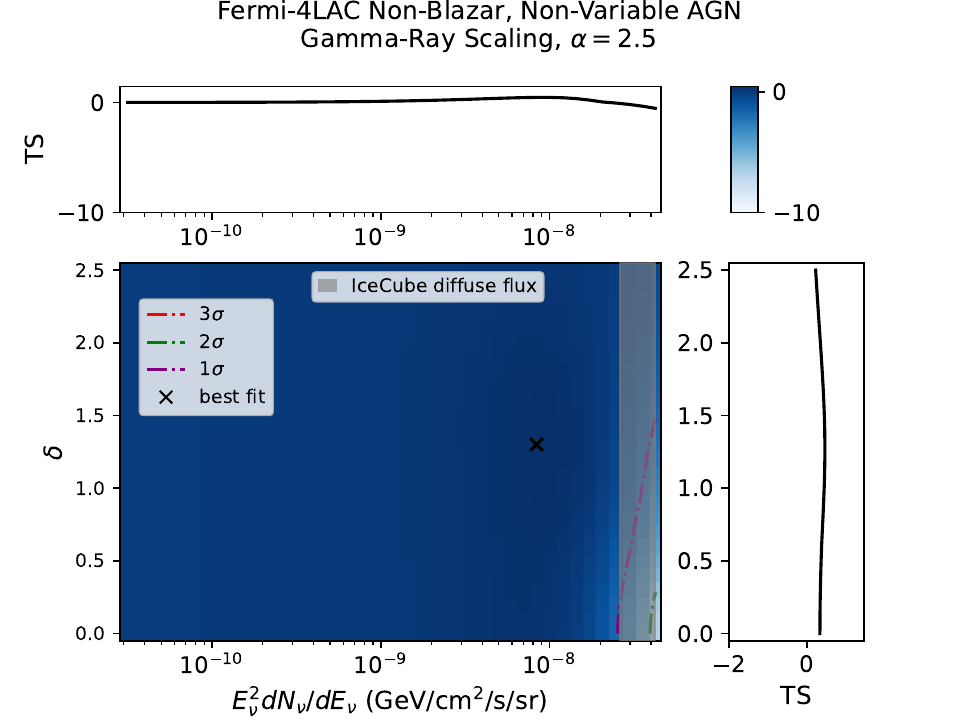}
\end{minipage}\hfill
\begin{minipage}{0.5\textwidth}
\centering
\includegraphics[width=1.0\linewidth]{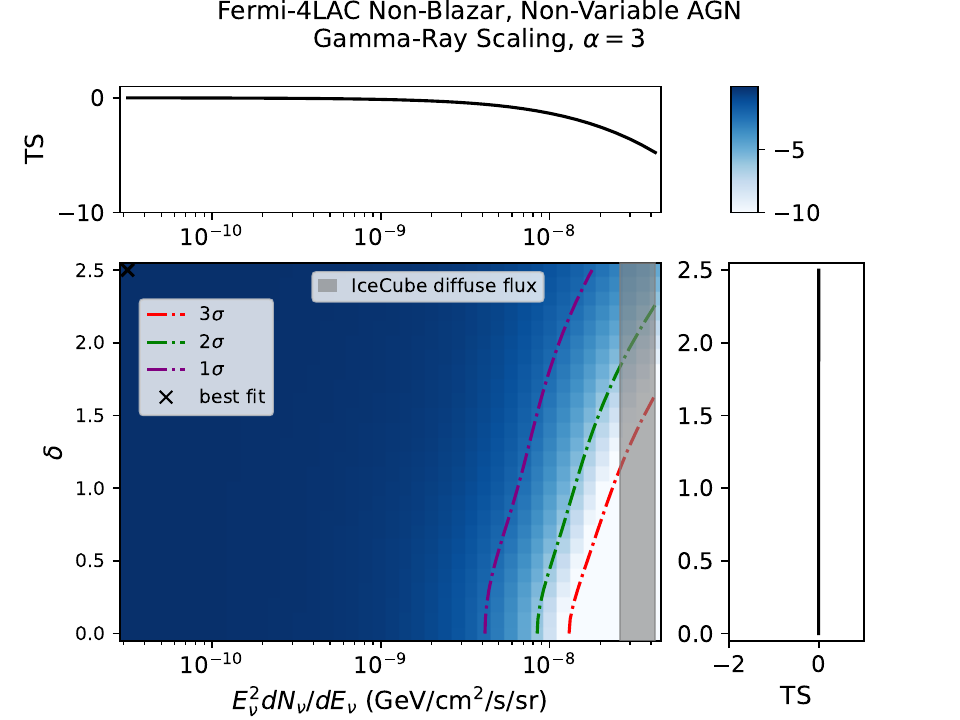}
\end{minipage}

\vspace{0.6cm}

\begin{minipage}{0.5\textwidth}
\centering
\includegraphics[width=1.0\linewidth]{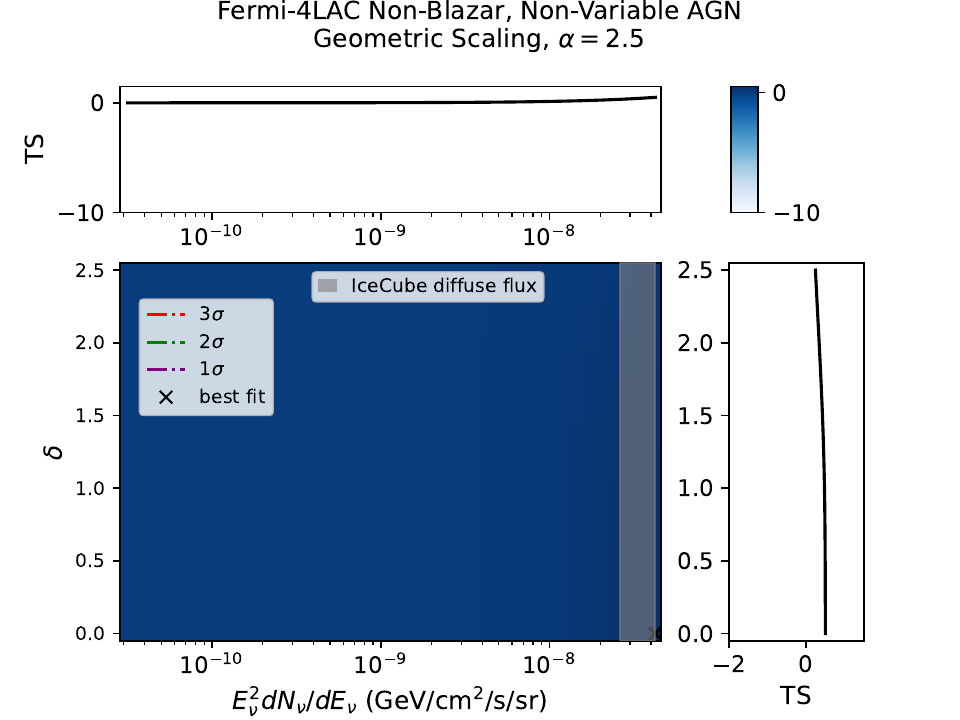}
\end{minipage}\hfill
\begin{minipage}{0.5\textwidth}
\centering
\includegraphics[width=1.0\linewidth]{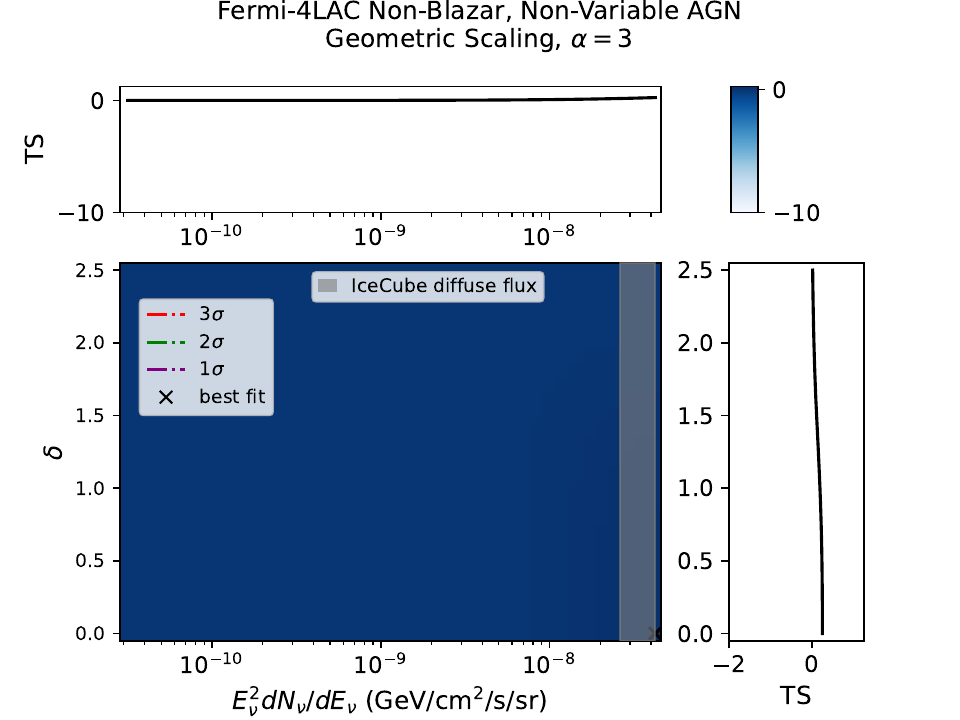}
\end{minipage}

\caption{As in Fig.~\ref{4LAC_all_nonblazars_fig}, but restricting our sample to non-variable, non-blazar gamma-ray-bright AGN.}
\label{4LAC_nonvar_nonblazars_fig}
\end{figure}

Finally, we consider the 64 non-blazar AGN contained in the Fermi 4LAC-DR3 catalog (of which 40 are non-variable). Our results in this case are shown in Fig.~\ref{4LAC_all_nonblazars_fig} and Fig.~\ref{4LAC_nonvar_nonblazars_fig} for the total and the non-variable populations, respectively. In each of the scenarios tested, we identify no statistically significant evidence of neutrino emission from this class of sources. Our constraints on the neutrino emission from this source population are summarized in Tables~\ref{4LAC_all_nonblazars_table} and~\ref{4LAC_nonvariable_nonblazars_table}.


\newpage

\clearpage

\vspace{1cm}

\clearpage

\begin{table}[t]
\centering
\textbf{Gamma-Ray-Bright, Blazar AGN (Fermi, 4LAC-DR3)} \\
\begin{tabular}{c|c|c}
Hypothesis & Significance~($\sigma$) & $2\sigma$ Limits on Contribution to $F^\text{diffuse}_{\nu}$  \\
\hline
Gamma-Ray Scaling, $\alpha=2.5$  & 0.0 & $F^{\rm AGN}_{\nu} < 7.0\%$  \\
\hline
Geometric Scaling, $\alpha=2.5$ & 0.0 & $F^{\rm AGN}_{\nu} < 14.5\%$ \\
\hline
Gamma-Ray Scaling, $\alpha=3.0$  & 0.0 & $F^{\rm AGN}_{\nu} < 3.2\%$  \\
\hline
Geometric Scaling, $\alpha=3.0$ & 0.0 &  $F^{\rm AGN}_{\nu} < 4.7\%$  \\
\end{tabular}




\caption{A summary of our results for gamma-ray-bright, blazar AGN. The constraints on the contribution to IceCube's diffuse neutrino flux are presented in terms of the flux from the entire source class, obtained after applying a modest completeness factor (see Table~\ref{completeness_factors_all}).
\label{4LAC_all_blazars_table}}
\end{table}

\vspace{1cm}

\begin{table}[H]
\centering
\textbf{Gamma-Ray-Bright, Non-Variable Blazar AGN (Fermi, 4LAC-DR3)} \\
\begin{tabular}{c|c|c}
Hypothesis & Significance~($\sigma$) & $2\sigma$ Limits on Contribution to $F^\text{diffuse}_{\nu}$  \\
\hline
Gamma-Ray Scaling, $\alpha=2.5$ & 0.0 &  $F^{\rm AGN}_{\nu} < 5.9\%$  \\
\hline
Geometric Scaling, $\alpha=2.5$ & 0.0 & $F^{\rm AGN}_{\nu} < 15.6\%$  \\
\hline
Gamma-Ray Scaling, $\alpha=3.0$ & 0.0 & $F^{\rm AGN}_{\nu} < 2.5\%$ \\
\hline
Geometric Scaling, $\alpha=3.0$ & 0.0  & $F^{\rm AGN}_{\nu} < 6.5\%$  \\
\end{tabular}

\caption{As in Table~\ref{4LAC_all_blazars_table}, but restricting our sample to non-variable gamma-ray-bright AGN.\label{4LAC_nonvar_blazars_table}}
\end{table}

\clearpage

\clearpage

\begin{table}[H]
\centering
\textbf{Gamma-Ray-Bright, Non-Blazar AGN (Fermi, 4LAC-DR3)} \\
\begin{tabular}{c|c|c}
Hypothesis & Significance~($\sigma$)  & Contribution to $F^\text{diffuse}_{\nu} \, (2\sigma)$   \\
\hline
Gamma-Ray Scaling, $\alpha=2.5$ & 0.0 & No Constraint  \\
\hline
Geometric Scaling, $\alpha=2.5$ & 0.0 & No Constraint \\
\hline
Gamma-Ray Scaling, $\alpha=3.0$ & 0.02 & No Constraint  \\
\hline
Geometric Scaling, $\alpha=3.0$ & 0.0 & No Constraint  \\
\end{tabular}

\caption{A summary of our results for gamma-ray-bright, non-blazar AGN. The constraints on the contribution to IceCube's diffuse neutrino flux are presented in terms of the flux from the entire source class, obtained after applying a modest completeness factor (see Table~\ref{completeness_factors_all}). \label{4LAC_all_nonblazars_table}}
\end{table}

\begin{table}[H]
\centering
\textbf{Gamma-ray-bright, Non-Variable, Non-Blazar AGN (Fermi, 4LAC-DR3)} \\
\begin{tabular}{c|c|c}
Hypothesis & Significance~($\sigma$) & Contribution to $F^\text{diffuse}_{\nu} \, (2\sigma)$   \\
\hline
Gamma-Ray Scaling, $\alpha=2.5$ & 0.26 & No Constraint  \\
\hline
Geometric Scaling, $\alpha=2.5$ & 0.28 & No Constraint \\
\hline
Gamma-Ray Scaling, $\alpha=3.0$ & 0.0 & No Constraint \\
\hline
Geometric Scaling, $\alpha=3.0$ & 0.15 & No Constraint  \\
\end{tabular}

\caption{As in Table~\ref{4LAC_all_nonblazars_table}, but restricting our sample to non-variable gamma-ray-bright non-blazar AGN. \label{4LAC_nonvariable_nonblazars_table}}
\end{table}

\subsection{Comparison with Other Work}

For recent and ongoing work on this subject by the IceCube Collaboration, see Ref.~\cite{IceCube:2025tmc} (for earlier work, see Refs.~\cite{IceCube:2016qvd,Smith:2020oac,Hooper:2018wyk}). Compared to the simpler procedure adopted in Refs.~\cite{IceCube:2016qvd,IceCube:2025tmc}, which constructs the signal hypothesis for the likelihood ratio according to $P(F_\nu|F_0)=\delta(F_\nu,F_0)$, where $F_0$ is the nominal neutrino flux hypothesized assuming proportionality to the electromagnetic flux, we allow for statistical scatter in the neutrino flux from each source, adopting $P(F_\nu|F_0)=(1/\sqrt{2\pi} \, \delta \, F_\nu)\, \exp(-\ln^2(F_\nu/F_0)/2\delta^2)$, where $\delta$ is a free parameter. While this sacrifices some sensitivity to a nominal ($\delta=0$) case due to the additional degree-of-freedom in the fit, it is more sensitive to cases in which the the gamma-ray and neutrino fluxes are correlated but not strictly proportional. The two stacking procedures have varying sensitivity to other hypotheses as well. For example, in a case where the fluxes are related by a scatter term of the form, $P(F_\nu,F_0)\propto \exp[-(F_\nu-F_0)^2/2\sigma_F^2]$, this should be more strongly constrained by the simpler model when $\sigma_F$ is small and by the source-fluctuation model when $\sigma_F$ is large. We further note that Ref.~\cite{IceCube:2025tmc} restricts their AGN sample to those in northern hemisphere in order to take advantage of the Northern Tracks IceCube dataset, which includes 13 years of data and an improved description of the point spread function, while this work includes AGN across the entire sky and uses the latest public IceCube dataset, consisting of 10 years of data. That work considers only AGN with a measured redshift and tests both a linear relationship between gamma-ray luminosity and neutrino output as well as a quadratic relationship between these quantities. This work, in contrast, tests the geometric scaling and linear relationship and requires sources to have a measured redshift only in the geometric case. This work also considers independently the full blazar and non-variable blazar samples, while Refs.~\cite{IceCube:2016qvd,IceCube:2025tmc} divided their AGN by classification and, in the case of  Ref.~\cite{IceCube:2016qvd}, by synchrotron peak.

\begin{table}[H]
\centering
\begin{tabular}{lcccc}
\toprule
Source & TS & $N_{\rm events}$ &$\alpha$ & $z$ \\
\midrule
NGC 1068                      & 28.11 & 63.4 & 3.0 & 0.0038                  \\
SWIFT J1041.4-1740       & 9.28 & 14.4 & 3.5 & 0.088         \\
SWIFT J0202.4+6824A/B        & 9.22 & 36.9 &2.5 & 0.0119                \\

SWIFTJ0744.0+2914                   & 9.08 & 29.6 &2.75 & 0.0160     \\
NGC 4151                       & 8.95 & 30.1 &2.5 & 0.0315                 \\
NGC 3079                   & 8.93 & 27.6 & 3.0 & 0.0037                 \\
SWIFT J1211.3-3935       & 7.73 & 6.5 &3.5 & 0.0228         \\
SWIFT J0350.1-5019       & 7.44 & 14.5 & 2.75 & 0.0359                \\
NGC 1194                   & 6.65 & 28.0 & 3.0 & 0.0136             \\
SWIFT J0741.4-5447                   & 6.30 & 7.7 & 3.5 & 0.713                   \\

\bottomrule
\end{tabular}
\caption{The Swift-BAT AGN for which we find evidence of neutrino emission at a level of ${\rm TS} > 6.18$ (corresponding to $2\sigma$ for 2 degrees-of-freedom). For each source, we provide the number of events and spectral index, $\alpha$, which yield the highest value of the test statistic (TS).}
\label{brightest}
\end{table}

\section{Discussion}
\label{discussion}

\begin{figure}[t]
\centering
\renewcommand{\arraystretch}{1.2}
\setlength{\tabcolsep}{-1pt}

\newcommand{\alpharow}[1]{\smash{\raisebox{2cm}[0pt][0pt]{$\alpha = #1$}}}


\makebox[\textwidth][l]{\hspace{-2.6cm}

\begin{tabular}{m{3cm} c c c}
   
    \centering  &
    \begin{subfigure}{0.31\textwidth}\includegraphics[width=\linewidth]{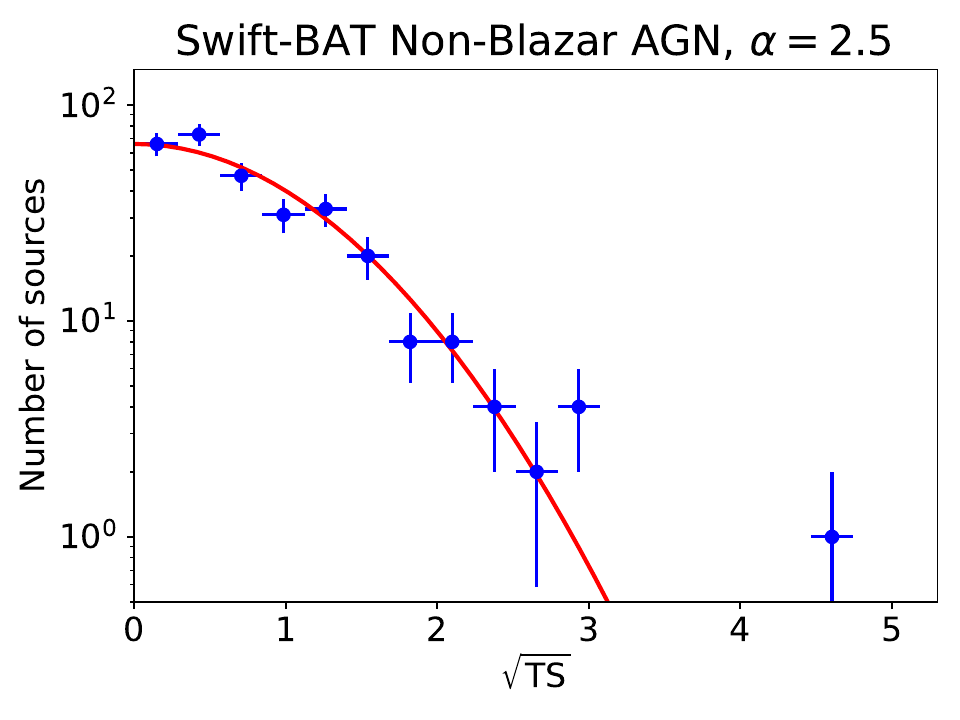}\end{subfigure} &
    \begin{subfigure}{0.31\textwidth}\includegraphics[width=\linewidth]{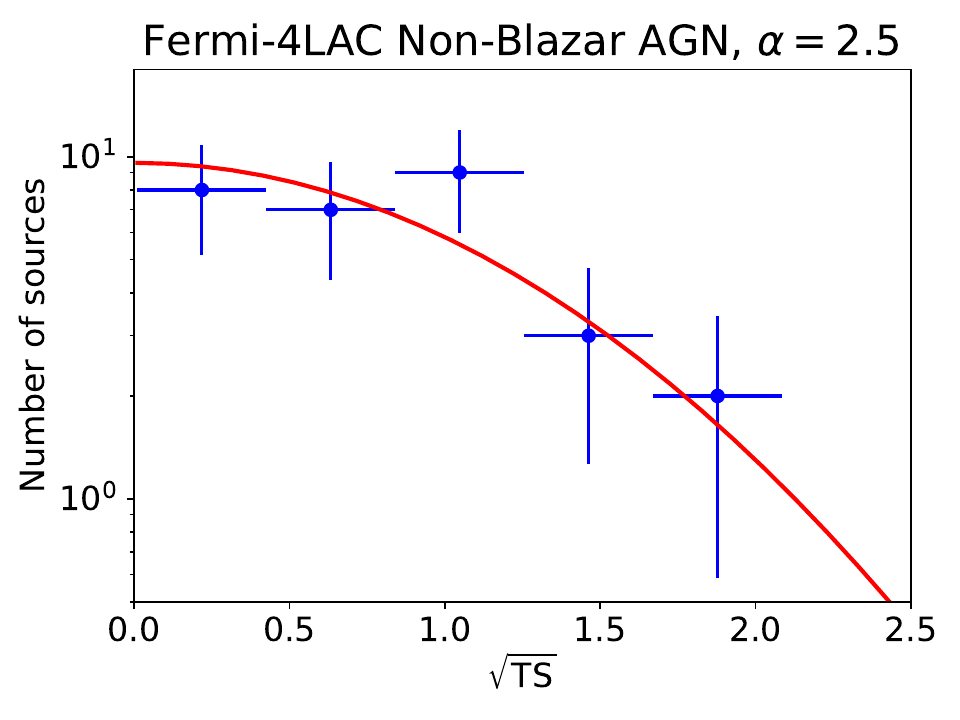}\end{subfigure} &
    \begin{subfigure}{0.31\textwidth}\includegraphics[width=\linewidth]{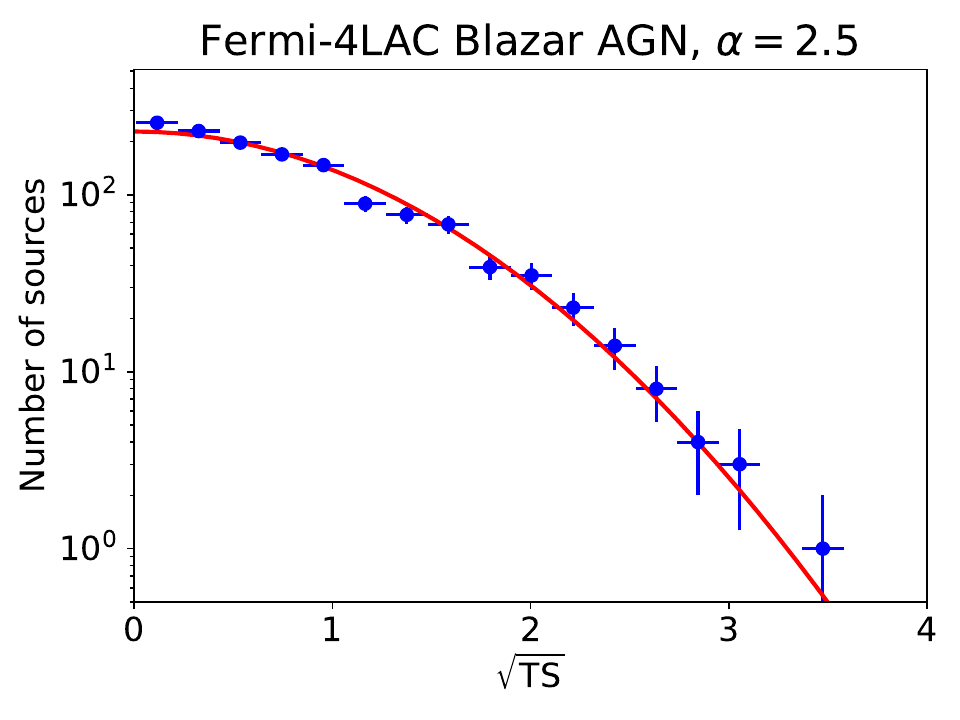}\end{subfigure} \\[5pt]

   \centering &
   \begin{subfigure}{0.31\textwidth}\includegraphics[width=\linewidth]{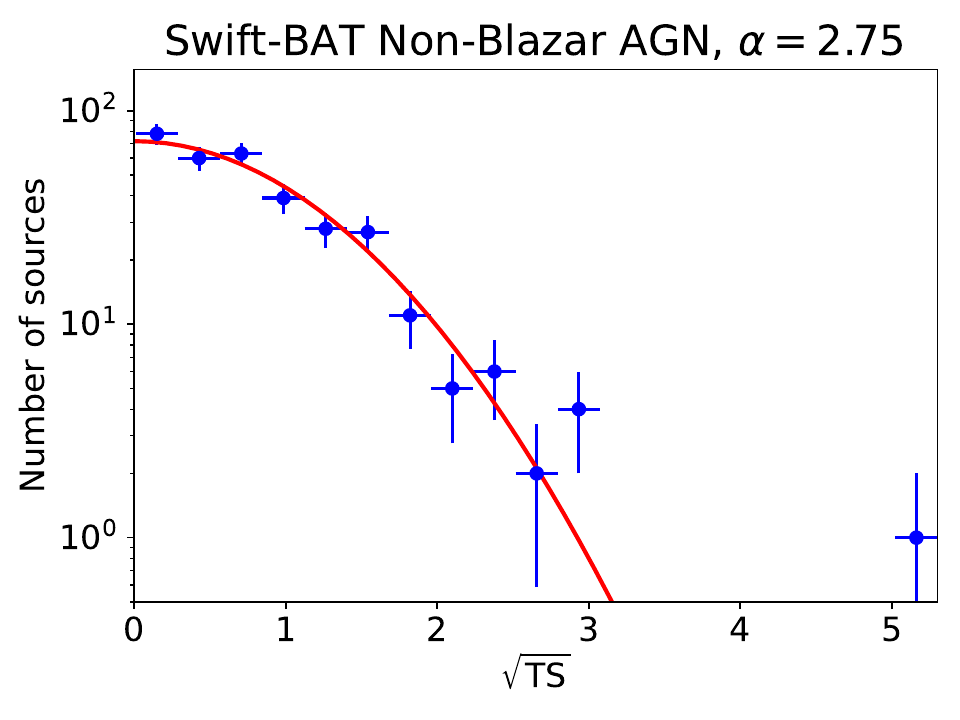}\end{subfigure} &
   \begin{subfigure}{0.31\textwidth}\includegraphics[width=\linewidth]{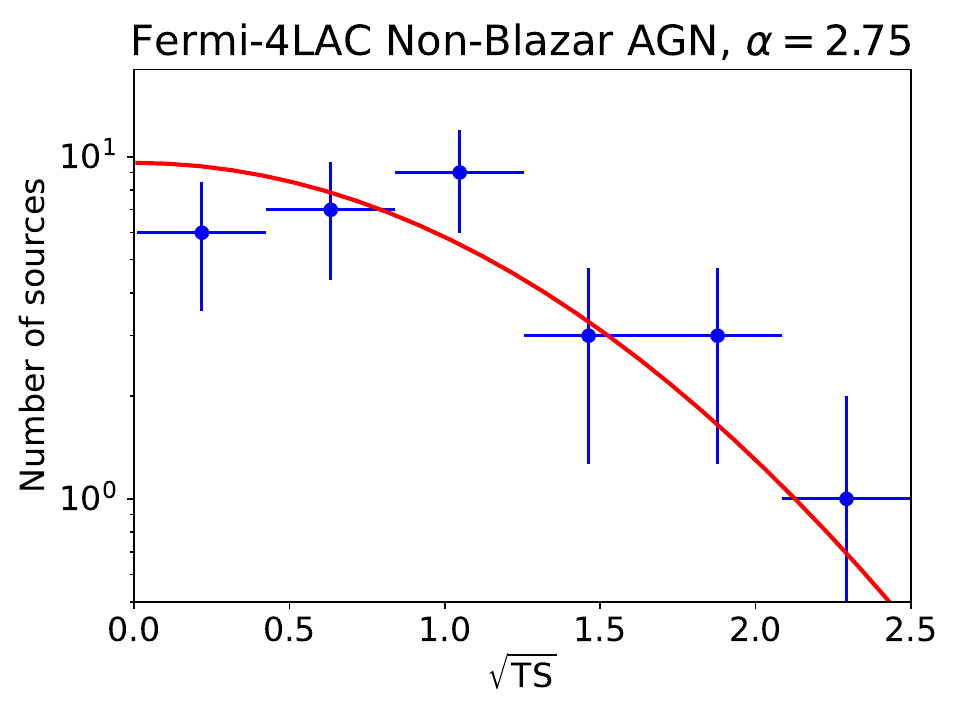}\end{subfigure} &
   \begin{subfigure}{0.31\textwidth}\includegraphics[width=\linewidth]{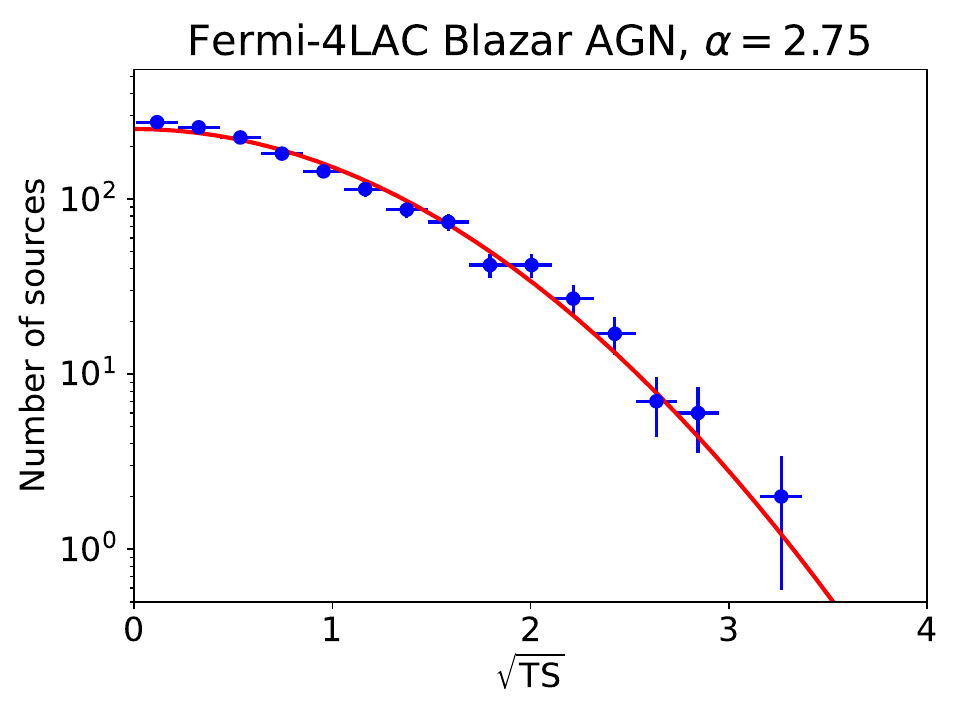}\end{subfigure} \\[5pt]

   \centering &
   \begin{subfigure}{0.31\textwidth}\includegraphics[width=\linewidth]{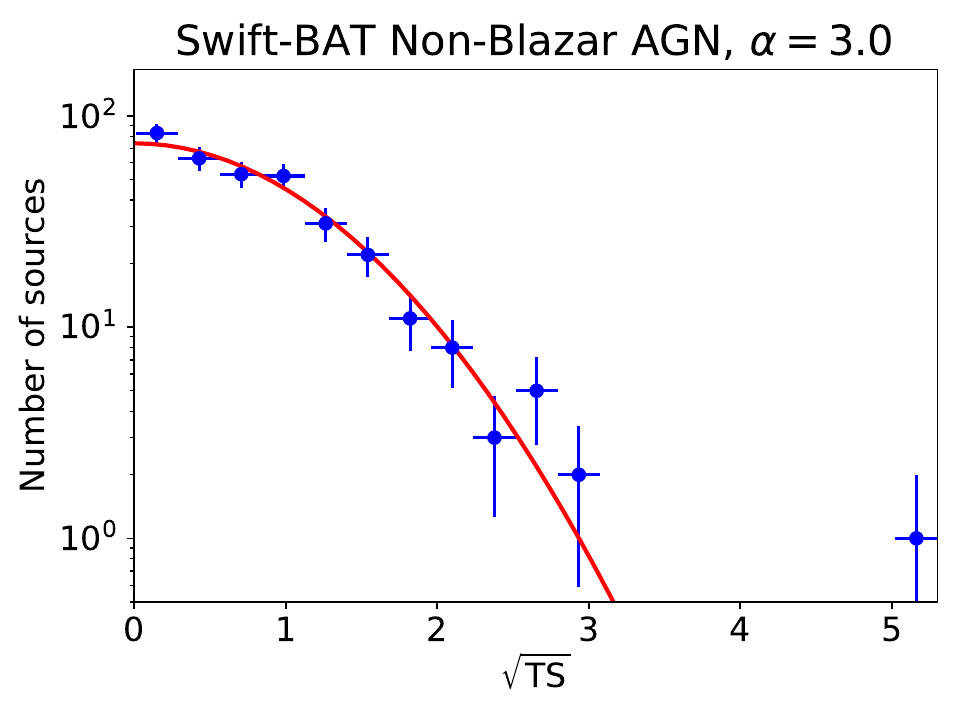}\end{subfigure} &
   \begin{subfigure}{0.31\textwidth}\includegraphics[width=\linewidth]{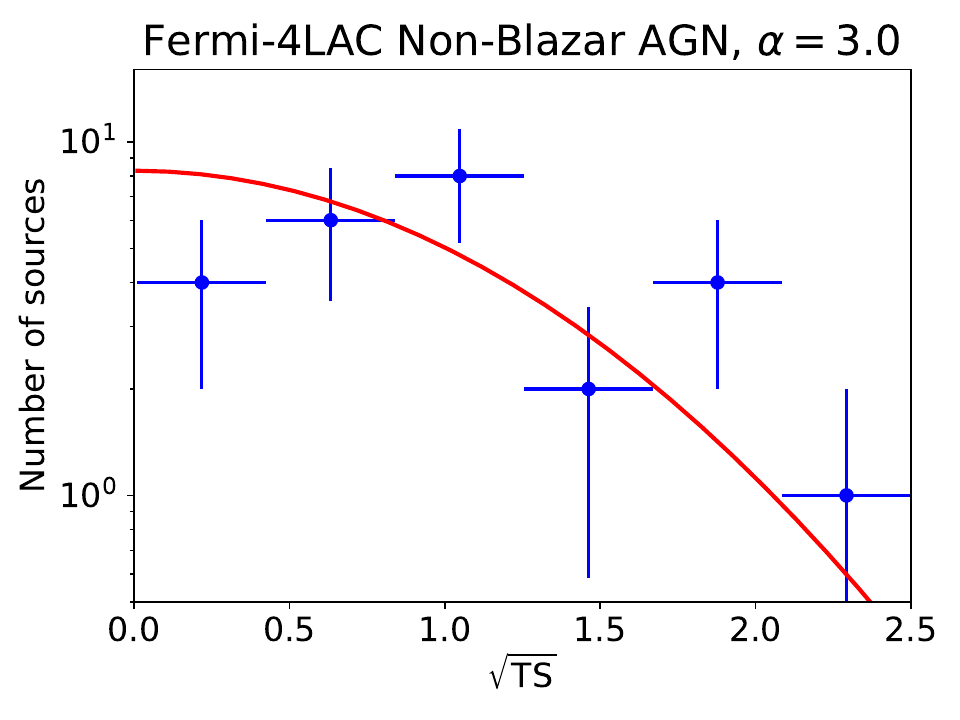}\end{subfigure} &
   \begin{subfigure}{0.31\textwidth}\includegraphics[width=\linewidth]{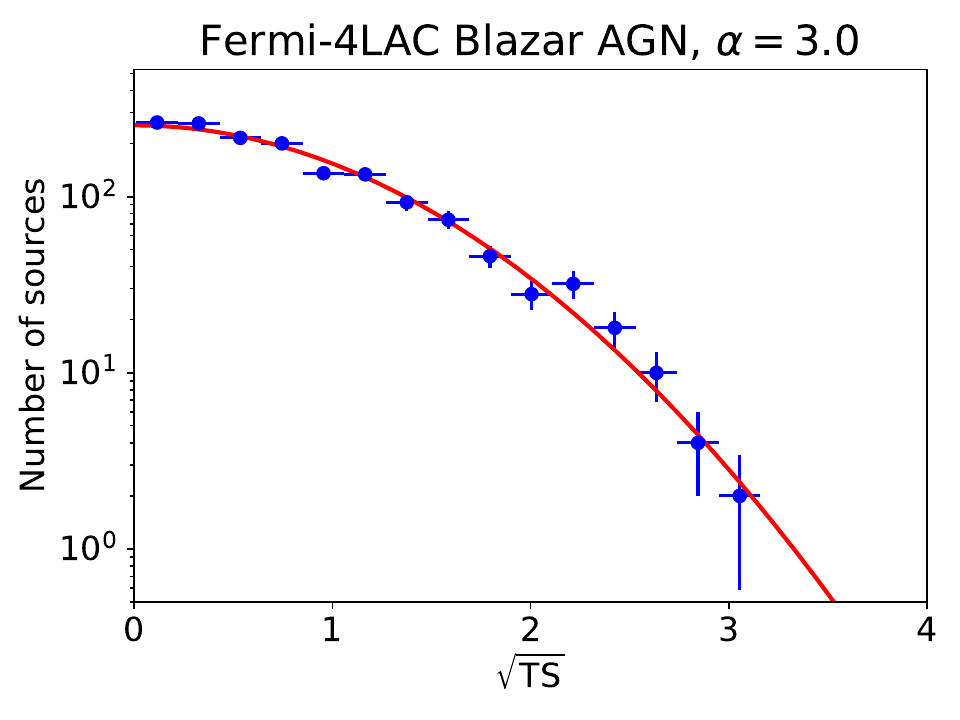}\end{subfigure} \\[5pt]

   \centering &
   \begin{subfigure}{0.31\textwidth}\includegraphics[width=\linewidth]{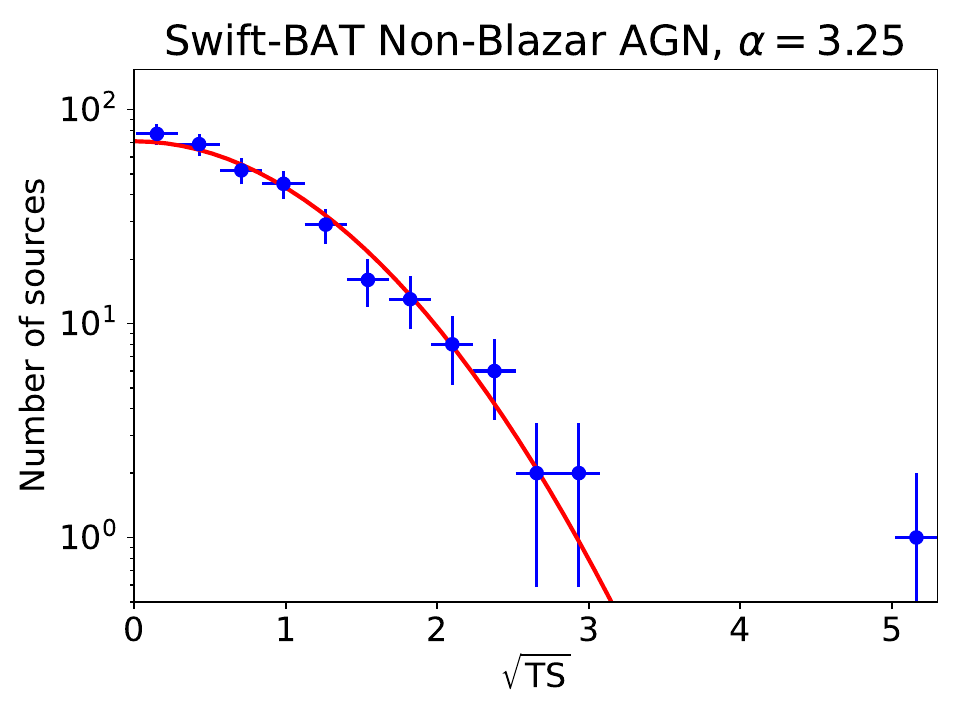}\end{subfigure} &
   \begin{subfigure}{0.31\textwidth}\includegraphics[width=\linewidth]{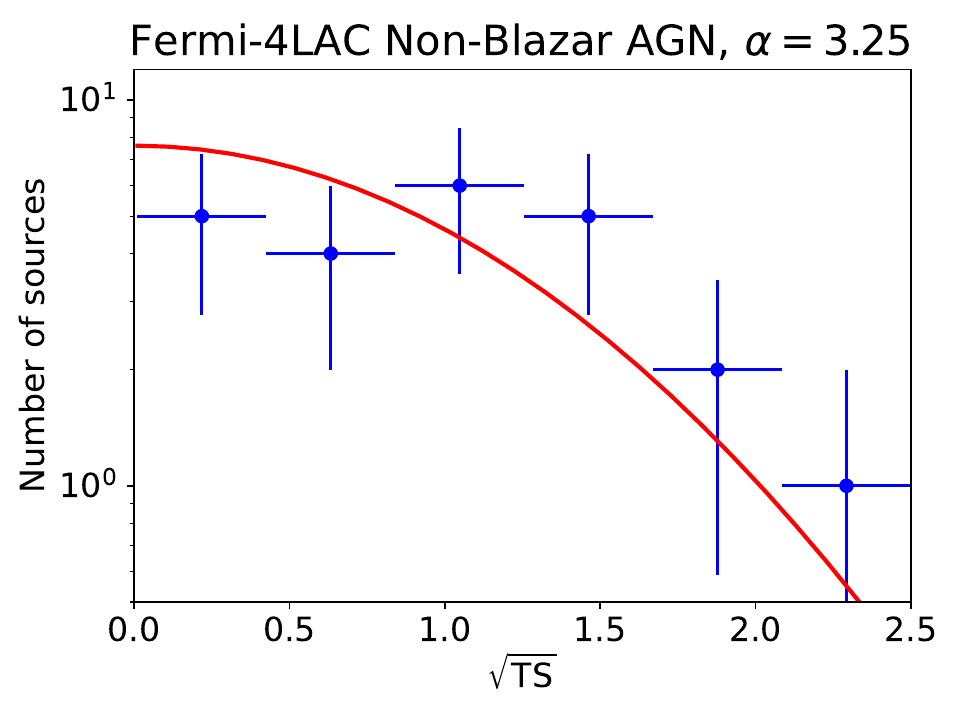}\end{subfigure} &
   \begin{subfigure}{0.31\textwidth}\includegraphics[width=\linewidth]{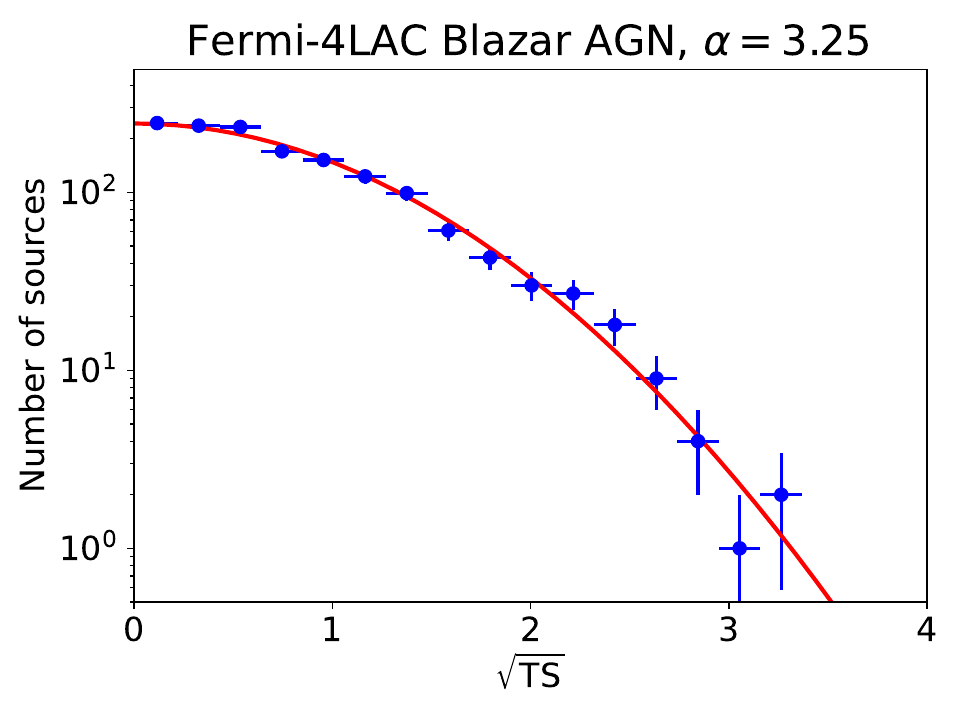}\end{subfigure} \\[5pt]

   \centering &
   \begin{subfigure}{0.31\textwidth}\includegraphics[width=\linewidth]{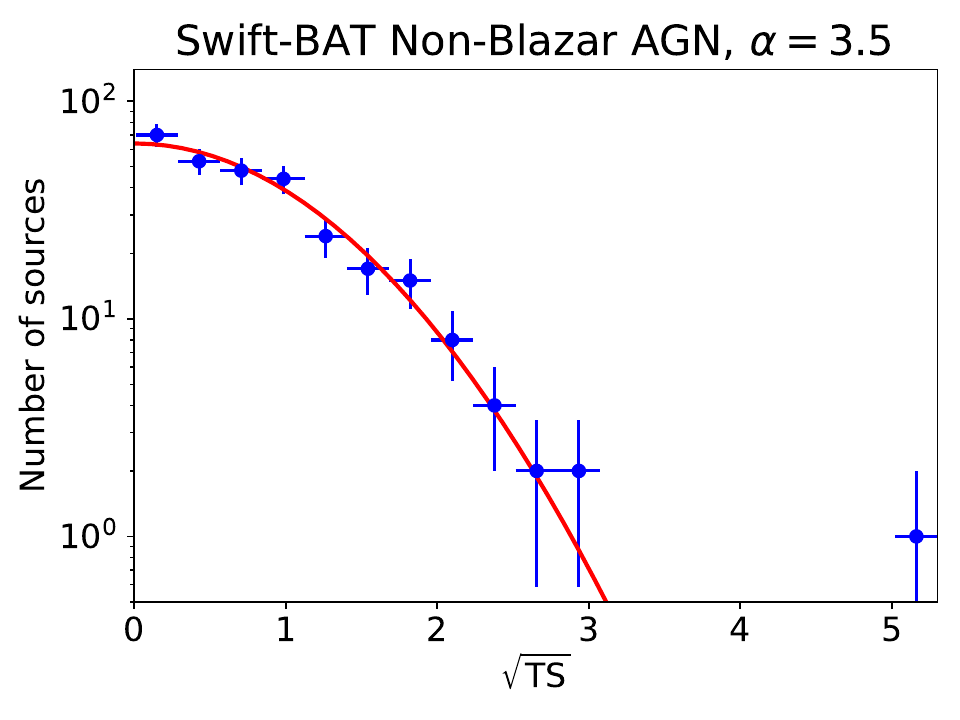}\end{subfigure} &
   \begin{subfigure}{0.31\textwidth}\includegraphics[width=\linewidth]{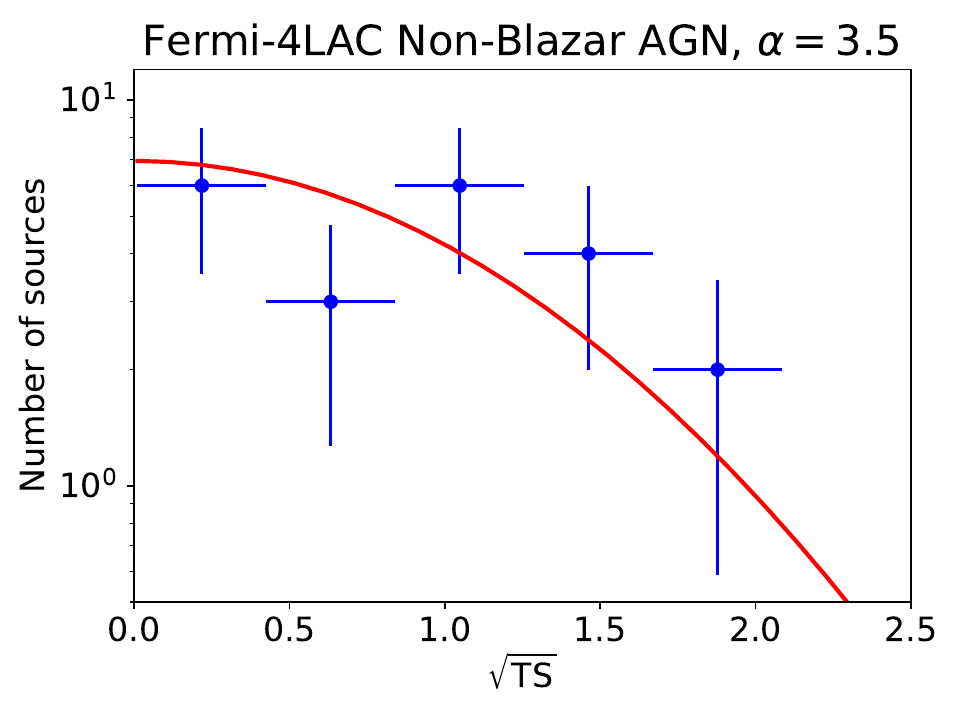}\end{subfigure} &
   \begin{subfigure}{0.31\textwidth}\includegraphics[width=\linewidth]{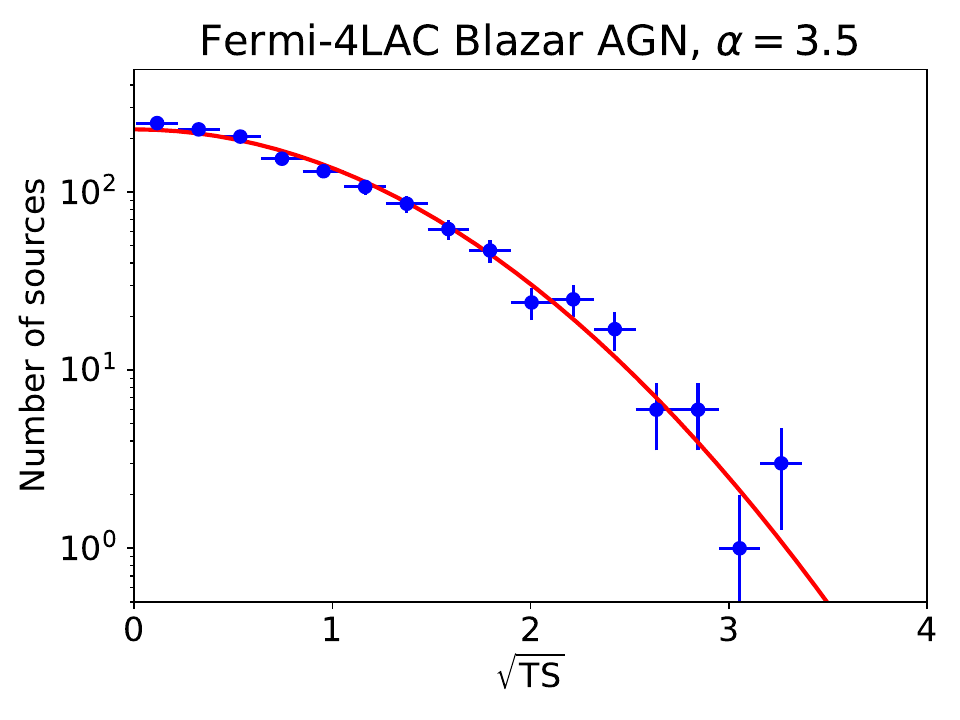}\end{subfigure} \\

\end{tabular}
}
\caption{The distribution of the test statistic (TS) for each AGN source sample and for five values of the neutrino spectral index, $\alpha$. With the exception of NGC 1068 (which is the $\sqrt{{\rm TS}}> 4-5$ outlier in each of the Swift-BAT frames), these distribution are consistent with gaussian variations, thereby demonstrating the lack of a detectable excess.\\ \\}
\label{fig:agn_grid}
\end{figure}

Most of the individual AGN considered in this analysis do not produce a detectable flux of high-energy neutrinos. In Fig.~\ref{fig:agn_grid}, we plot the distribution of the TS values found for of the each source populations considered and for five values of the neutrino spectral index, $\alpha$. In each case, the distribution is consistent with gaussian variations, thereby demonstrating the lack of any detectable excess. The exception to this is the case of X-ray-bright AGN, for which NGC 1068 is a clear outlier from the gaussian prediction, as seen in Table~\ref{table_swift_full}.  


In Table~\ref{brightest}, we provide a list of the Swift-BAT sources which yield the highest TS in our analysis. This table includes all such sources with ${\rm TS} > 6.18$, corresponding to $2\sigma$ (pre-trials) for 2 degrees-of-freedom. Note that given the number of sources contained in this catalog, this number of $>2\sigma$ sources is consistent with that expected from statistical fluctuations. That said, the fact that the highest TS sources are overwhelmingly X-ray-bright and at low redshift suggests that some of these sources may be authentic neutrino emitters; see also Refs.~\cite{icecube_ngc4151,Abbasi:2025tas}.

Lastly, we note that our best fit to the Swift-BAT sample of non-blazar AGN corresponds to a scenario in which the neutrino flux from each of these sources is, on average, proportional to its measured X-ray flux. As can be seen in Fig.~\ref{swift_full}, however, these fits also prefer a relatively large degree of source-to-source variation in this relationship, $\delta \sim 1.7$. The source NGC 1068 is, in fact, a $\sim 2 \sigma$ outlier among this source population, producing $e^{2\delta} \sim 30$ times more neutrino emission than would be predicted from a more typical source with the same X-ray flux.

\section{Conclusions}
\label{conclusions}

In this study, we have analyzed 10 years of publicly available IceCube data in an effort to measure or constrain the high-energy neutrino emission from known populations of AGN. We have tested scenarios in which AGN produce neutrino emission at a level proportional to their observed X-ray or gamma-ray emission, or in proportion to the inverse square of their luminosity distance. We have allowed for source-to-source variation in this proportionality and have considered separately sources which do and do not exhibit variability in their observed emission. 

Our main conclusions are summarized as follows:
\begin{itemize}
\item{As in earlier studies~\cite{IceCube:2016qvd,Smith:2020oac,Hooper:2018wyk}, we find no evidence of neutrino emission from gamma-ray-bright blazars. These sources can produce no more than $15.6\%$ of the diffuse neutrino flux observed by IceCube.}
\item{We find little evidence ($<0.28\sigma$) for neutrino emission from gamma-ray-bright, non-blazar AGN. We cannot, however, rule out the possibility that this class of sources contributes significantly to IceCube's diffuse flux.}
\item{We find significant evidence ($3.15-4.18\sigma$) of neutrino emission from the X-ray-bright, non-blazar AGN contained in the Swift-BAT catalog (see also, Refs.~\cite{IceCube:2024dou,IceCube:2024ayt}). Much of this statistical significance, however, comes from a single source, NGC 1068. If NGC 1068 is excluded from our analysis, the statistical significance of neutrino emission from this source class falls to 1.35$\sigma$.}
\item{In addition to NGC 1068, we identify evidence of neutrino emission from several other nearby, X-ray-bright Seyfert galaxies, including SWIFT J1041.4-1740 ($2.6\sigma$), SWIFT J0744.0+2914 ($2.6\sigma$), J0202.4+6824A/B ($2.6\sigma$), NGC 4151 ($2.5\sigma$), and NGC 3079 ($2.5\sigma$). Although the significance of these signals does not overcome the trials factor associated with our search, the fact that these high TS sources are each nearby and X-ray bright suggests that this source population could very well contain a significant number of high-energy neutrino emitters, many of which would likely lie within the discovery reach of future large-volume neutrino telescopes~\cite{IceCube-Gen2:2020qha,P-ONE:2020ljt,TRIDENT:2022hql,Huang:2023mzt}.}
\item{Our best fits correspond to scenarios in which the neutrino emission from a given AGN is, on average, proportional to its X-ray flux, but with significant source-to-source variation, $\delta \sim 1.7$. Our analysis further suggests that between 11.2\% and 100\% of IceCube's diffuse neutrino flux arises from X-ray-bright AGN.}
\item{With the exception of NGC 1068 and other X-ray-bright AGN, no known class of extragalactic sources has been found to be correlated with the neutrino events observed by IceCube. This indicates that the diffuse, high-energy neutrino flux must originate from a large number of relatively faint sources. As stated above, these sources could consist largely of non-blazar AGN. Alternatively, starburst or starforming galaxies could plausibly produce a significant fraction of IceCube's diffuse flux.}
\end{itemize}	

These results strengthen the case for scenarios in which a significant fraction of IceCube's diffuse neutrino flux is produced by non-blazar AGN, including X-ray-bright, gamma-ray-obscured Seyfert galaxies, such as NGC 1068, NGC 4151, and NGC 3079. This further suggests that a significant number of AGN could be identified as neutrino point sources by enhancing IceCube's performance and by by constructing future, large-volume neutrino detectors, such as IceCube-Gen2~\cite{IceCube-Gen2:2020qha,P-ONE:2020ljt,TRIDENT:2022hql,Huang:2023mzt}.

\section*{Acknowledgements}

We would like to thank Alicia Mand, Lu Lu, Matthias Thiesmeyer, Maxwell Nakos, Sam Hori, and Justin Vandenbroucke for their helpful suggestions. This work has been supported by the Office of the Vice Chancellor for Research at the University of Wisconsin-Madison, with
funding from the Wisconsin Alumni Research Foundation and in part by the U.S. National Science Foundation under grants~PHY-2209445 and OPP-2042807. 


\bibliographystyle{JHEP}
\bibliography{biblio.bib}


%

\appendix 



\section{Muon Effective Area}
\label{muon_area}

The IceCube 10-year public data release includes the detector's muon neutrino effective area for each of the detector's configurations, as a function of neutrino energy and declination, $A_\nu (E_\nu, \delta)$. This area can be combined with the neutrino spectrum from a given source to calculate the mean number of neutrino-induced muon track events expected from that source per unit time,
\begin{equation}
    N_{\rm events} = \int \frac{dN_{\nu}}{dE_\nu}(E_{\nu}) \,  A_\nu(E_\nu, \delta) \, dE_\nu \,.
    \label{n_neutrino_area}
\end{equation}
The publicly available IceCube dataset reports a value of the muon energy for each event (evaluated at the track's point of closest approach to the center of the detector). So although the neutrino effective area can be used to calculate the expected number of events, this alone cannot be straightforwardly used to predict the distribution of the muon energies. 

The differential rate of such events can also be represented in terms of the differential cross section for charged-current neutrino-nucleon scattering~\cite{UHE_interactions}, the muon energy loss rate, and the detector's muon effective area,
\begin{equation}
    \frac{dN_{\rm events}}{dE_{\mu}^f}= N_A\int_{E_{\nu}^{\rm min}}^{E_{\nu}^{\rm max}} \int_0^{E_{\nu}} dE_\nu \, dE_{\mu}^i  \,   \frac{dN_{\nu}}{dE_\nu}(E_{\nu}) \,\frac{d\sigma_{\nu N}^{\rm CC}}{dE_{\mu}^i}(E_\nu, E_{\mu}^i) \, \frac{dX_{\mu}}{dE_{\mu}}(E_{\mu}^f, E_{\mu}^i,\delta)  \, A_\mu(E_{\mu}^f,\delta)\, S(E_\nu) \, , 
    \label{n_muon_area}
\end{equation}
where $N_A$ is the number density of nucleons in ice, $E_{\mu}^i$ is the energy of the muon at the location of the neutrino scattering event, $E_{\mu}^f$ is the energy of the muon at its point of closest approach to the center of the detector, $\sigma_{\nu N}^{\rm CC}$ is the charged-current neutrino-nucleon cross section, $S(E_\nu)$ is an absorption factor which accounts for neutrino scattering in the Earth, and $A_{\mu}$ is the muon effective area of the detector. The quantity $dX_{\mu}/dE_{\mu}$ is equal to the distance traveled per unit energy loss of the muon in the detector, which is given by
\begin{align}
\frac{dX_{\mu}}{dE_{\mu}}(E_{\mu}^f, E_{\mu}^i,\delta) = \frac{1}{\alpha + \beta E_{\mu}^f},
\label{eq:dXdE}
\end{align}
where $\alpha \approx 2.4 \times 10^{-3} \, {\rm GeV/cm}$ and $\beta \approx 3 \times 10^{-6} \, {\rm cm}^{-1}$~\cite{Dutta:2000hh}. Instead of using the expression in Eq.~\ref{eq:dXdE}, $dX_{\mu}/dE_{\mu}$ should be replaced with zero if the following condition is met:
\begin{align}
D(\delta) < R_{\mu}(E_{\mu}^i) = \frac{1}{\beta} \ln \bigg[\frac{\alpha+\beta E_{\mu}^i}{\alpha+\beta E_{\mu}^f}\bigg],
\end{align}
where $D(\delta)$ is the distance from the center of the detector to the surface of the Earth along the incoming neutrino direction. 

To determine the muon effective area of IceCube as a function of $E_{\mu}^i$ and $\delta$, we integrate Eq.~\ref{n_muon_area} with respect to the final muon energy and equate the result with the rate given in Eq.~\ref{n_neutrino_area}. For each value of $\delta$ and $E_{\nu}$, this leads to 
\begin{equation}
  A_{\nu}(E_\nu,\delta)= 
  \int_0^{E_{\nu}}  \int_0^{E_{\mu}^i} dE_{\mu}^f  \, dE_{\mu}^i  \,\frac{d\sigma_{\nu N}^{\rm CC}}{dE_{\mu}^i}(E_\nu, E_{\mu}^i) \, \frac{dX_{\mu}}{dE_{\mu}}(E_{\mu}^f, E_{\mu}^i,\delta)  \, A_\mu(E_{\mu}^f,\delta)\, S(E_\nu) \, ,
\end{equation}
which can be solved to obtain the detector's muon effective area, $A_{\mu}(E_{\mu}^f,\delta)$. From this, we can use Eq.~\ref{n_muon_area} to calculate the rate of events per unit muon energy (at the track's point of closest approach to the center of the detector), which we can directly compare to the public IceCube dataset.

\section{Completeness Factors}
\label{completeness_factors}

For each source population and scaling hypothesis, we define the completeness factor as the ratio of the neutrino signal expected from an entire source class (including those sources not contained in the catalog) to that from the sum of the sources in the catalog. In this Appendix, we describe how these completeness factors were calculated. 

Starting with the case of the X-ray scaling hypothesis, for a source with an intrinsic X-ray flux, $F^{\rm intr}_X$, the predicted neutrino flux (evaluated at 30 TeV) is given by 
\begin{equation}
    F^{\rm obs}_{\nu} = c \, F^{\rm intr}_X \, (1+z)^{-\alpha}\, ,
\end{equation}
where $\alpha$ is the spectral index of the neutrino spectrum and $c$ is a proportionality constant. The total neutrino flux predicted from the entire source class can be written as
\begin{align}
    \label{total_fnu}
    F_{\nu}^{\text{total}} =& \int dz \, dL \,  \, \frac{dN}{dLdV_c} \, \frac{L}{4\pi D_L^2} \frac{dV_c}{dz}  \, (1+z)^{-\alpha} \\
    =& \int dz \, dL \,  \, \frac{dN}{dLdV_c} \, \frac{L\,c}{H(z)(1+z)^2}  \, (1+z)^{-\alpha},\nonumber 
\end{align}
where $dN/dL dV_c$ is the the number of sources per unit X-ray luminosity and comoving volume, $D_L$ is the luminosity distance, $dV_c/dz$ is the comoving volume per unit redshift, and the Hubble rate is given by
    $H(z) = H_0 \sqrt{\Omega_M(1+z)^3 +\Omega_\Lambda}$.
%
Throughout this study, we adopt the best-fit cosmological parameters, $H_0$, $\Omega_M$, and $\Omega_{\Lambda}$, as reported by the Planck Collaboration~\cite{planck2018}. 

To obtain the completeness factor, we divide the quantity given in Eq.~\ref{total_fnu} by the sum of the neutrino fluxes predicted from the members of the source catalog,
\begin{equation}
    C_f = \frac{F_{\nu}^{\text{total}}}{\sum_i F_{\nu, i}^{\text{obs}}}.
\end{equation}
For X-ray-bright, non-blazar AGN, we use the X-ray luminosity function and redshift distribution presented in Ref.\cite{nonblazar_xlf}. 


We apply the same procedure to calculate the completeness factors for those cases involving the gamma-ray scaling hypothesis. For gamma-ray-bright blazars, we use the gamma-ray luminosity function and redshift distribution described in Ref.~\cite{blazar_lf}. For 4LAC non-blazar AGN, completeness factors for the gamma-ray scaling were previously computed in Ref.~\cite{Smith:2020oac}, using the original data release of the 4LAC catalog. To update this, we have rescaled the corresponding completeness factors by the ratio of total gamma-ray flux from the members of the 4LAC-DR3 and 4LAC-DR1 catalogs. 

For the geometric scaling hypothesis, we calculate the completeness factors following the same procedure, but adopting a delta function for the luminosity function. Our completeness factors for the Swift-BAT and 4LAC-DR3 catalogs are given in Table~\ref{completeness_factors_all}.

\begin{table}[!t]

\centering
\vspace{0.5cm}
\renewcommand{\arraystretch}{1.3}
\setlength{\tabcolsep}{6pt}

\begin{tabular}{|l|c|c|c|c|c|}
\hline
\multirow{2}{*}{Swift Source Class} &
\multicolumn{2}{|c|}{Hard X-Ray} &
\multicolumn{2}{|c|}{Soft X-Ray} &
\multicolumn{1}{|c|}{Geometric}\\
\hhline{|~|-----|}
 & $\alpha = 2.5$ & $\alpha = 3.0$
 & $\alpha = 2.5$ & $\alpha = 3.0$
 & $\alpha = 2.5, 3.0$  \\
\hline
All Non-Blazar AGN & 24.3 & 20.7 & 30.0 & 25.4 & 11.1   \\
\hline
\makecell{Non-Blazar AGN\,\,\,\,\,\,\,\,\,\,\,\,\,\,\,\,\,\, \\(Excluding NGC 1068)\,\,\,\,} & 24.5 & 20.9 & 30.8 & 26.1 & 11.1   \\
\hline
\end{tabular}
\centering
\vspace{0.5cm}
\renewcommand{\arraystretch}{1.3}
\setlength{\tabcolsep}{6pt}

\begin{tabular}{|l|c|c|}
\hline
\multirow{2}{*}{Fermi Source Class} & 
\multicolumn{1}{|c|}{Gamma Ray} & 
\multicolumn{1}{|c|}{Geometric}   \\
\hhline{|~|--|}
 & $\alpha = 2.5, 3.0$ 
 & $\alpha = 2.5, 3.0$  \\
\hline
All Blazars & 1.12 & 1.49  \\
\hline
Non-Variable Blazars & 1.12 & 1.49   \\
\hline
All Non-Blazar AGN & 48.5 & 2710  \\
\hline
Non-Variable, Non-Blazar AGN & 360 & 2710 \\
\hline
\end{tabular}

\caption{Completeness factors for different source classes and hypotheses for the Swift-BAT and 4LAC-DR3 populations.}
\label{completeness_factors_all}
\end{table}

\end{document}